\documentclass{scrartcl}

\usepackage{authblk}

\usepackage{tikz}
\usepackage{pgf}
\usepackage{pgfplots}
\usepackage{pgfplotstable}
\usepackage{enumitem}
\usepackage{subcaption}

\usepackage{amsmath}
\usepackage{amsfonts}

\usepackage{booktabs}

\usepackage{hyperref}
\usepackage{cleveref}

\usepackage[backend=biber,sorting=nyvt]{biblatex}
\usetikzlibrary{calc, intersections, positioning, hobby, fit, shapes.misc, patterns}

\makeatletter
\pgfkeys{%
  /tikz/on layer/.code={
    \pgfonlayer{#1}\begingroup
    \aftergroup\endpgfonlayer
    \aftergroup\endgroup
  },
  /tikz/node on layer/.code={
    \gdef\node@@on@layer{%
      \setbox\tikz@tempbox=\hbox\bgroup\pgfonlayer{#1}\unhbox\tikz@tempbox\endpgfonlayer\egroup}
    \aftergroup\node@on@layer
  },
  /tikz/end node on layer/.code={
    \endpgfonlayer\endgroup\endgroup
  }
}
\def\node@on@layer{\aftergroup\node@@on@layer}
\makeatother

\pgfplotsset{select coords between index/.style 2 args={
    x filter/.code={
        \ifnum\coordindex<#1\fi
        \ifnum\coordindex>#2\fi
    }
}}

\pgfkeys{/pgfplots/tuftelike/.style={
  semithick,
  tick style={major tick length=4pt,semithick,black},
  separate axis lines,
  axis x line*=bottom,
  axis x line shift=10pt,
  axis y line*=left,
  axis y line shift=10pt,
  enlarge x limits=false,
  enlarge y limits=false
  }}

\pgfdeclarelayer{bbg}    %
\pgfdeclarelayer{bg}    %
\pgfdeclarelayer{fg}    %
\pgfsetlayers{bbg,bg,main,fg}  %

\definecolor{pblue}{RGB}{0,102,189}  %
\definecolor{porange}{RGB}{243,98,33}  %

\tikzstyle{sample}=[circle,draw,inner sep=1.5pt,thick,fill=white]
\tikzstyle{stample}=[circle,inner sep=1.5pt,fill]

\tikzset{>=latex}

\usepackage[frozencache]{minted}

\makeatletter
\newcommand{\minted@style@bw}{%
  \renewcommand\fcolorbox[3][]{##3}%
  \renewcommand\textcolor[3][]{##3}%
}

\newcommand{\cppinline}[1]{\mintinline[style=sas]{C++}{#1}}
\newcommand{\xmlinline}[1]{\mintinline[style=sas]{xml}{#1}}
\newcommand{\pyinline}[1]{\mintinline[style=sas]{python}{#1}}

\usepackage{multirow}

\newcolumntype{L}[1]{>{\raggedright\let\newline\\\arraybackslash\hspace{0pt}}m{#1}}
\newcolumntype{C}[1]{>{\centering\let\newline\\\arraybackslash\hspace{0pt}}m{#1}}
\newcolumntype{R}[1]{>{\raggedleft\let\newline\\\arraybackslash\hspace{0pt}}m{#1}}

\title{A waveform iteration implementation for black-box multi-rate higher-order coupling}

\author[1]{Benjamin Rodenberg}
\affil[1]{TUM School of Computation, Information and Technology, Technical University of Munich, \texttt{benjamin.rodenberg@cit.tum.de}}
\author[2]{Benjamin Uekermann}
\affil[2]{Institute for Parallel and Distributed Systems (IPVS), University of Stuttgart, \texttt{benjamin.uekermann@ipvs.uni-stuttgart.de}}

\ifpdf
\hypersetup{
  pdftitle={A waveform iteration implementation for black-box multi-rate higher-order coupling},
  pdfauthor={Benjamin Rodenberg, Benjamin Uekermann}
}
\fi

\begin{document}

\maketitle

\begin{abstract}
Many multiphysics simulations involve processes evolving on disparate time scales, posing a challenge for efficient coupling. A naive approach that synchronizes all processes using the smallest time scale wastes computational resources on slower processes and typically achieves only linear convergence in time. Waveform iteration is a promising numerical technique that enables higher-order, multi-rate coupling while treating coupled components as black boxes. However, applying this approach to PDE-based coupled simulations is nontrivial.

In this paper, we integrate waveform iteration into the black-box coupling library preCICE with minimal modifications to its API. We detail how this extension interacts with key preCICE features, including data mapping for non-matching meshes, quasi-Newton acceleration for strongly coupled problems, and parallel peer-to-peer communication. We then showcase that waveform iteration significantly reduces numerical errors---often by orders of magnitude. This advancement greatly enhances preCICE, benefiting its extensive user community.\footnote{
The manuscript summarizes key parts of the dissertation of Rodenberg \cite{Rodenberg2025}.}
\end{abstract}

\section{Introduction}

Multiphysics simulations present numerous challenges. Among them, the combination of complex components and their efficient coupling in the time domain are two key issues explored in this paper. The first challenge can often be addressed through black-box coupling, which relies on minimal information exchange between components. However, the second challenge requires making the most of this limited information. In the time domain, we must account for components whose underlying processes evolve on disparate time scales. Furthermore, different components may employ distinct time integration methods with varying mathematical properties.

These challenges are common across many application domains. For instance, Earth system modeling necessitates coupling physical processes that evolve over vastly different time scales \cite{Gross2018}. The dynamics of ice sheets, for example, are affected by subglacial hydrological systems that evolve slowly in the interior but experience abrupt changes at the margins during peak melt seasons \cite{Abele2025}. More broadly, the dynamics of ice sheets, oceans, and the atmosphere all operate on different temporal scales.
Similar challenges arise in mechanical engineering. Munafò et al., for example, study inductively coupled plasma wind tunnels that combine fluid, electromagnetic, and solid models, whose time scales differ by two to three orders of magnitude \cite{Munafo2022}. Even classical coupled problems, such as fluid-structure interaction (FSI)---whether in hemodynamics or aeroelasticity---face challenges in the time domain. While these may not be as extreme, using different time steps and integration methods can still be advantageous to efficiently satisfy stability constraints \cite{Kersschot2021}.

\begin{figure}
\center{
  \begin{subfigure}[b]{0.49\textwidth}
    \centering
        \begin{tikzpicture}
    \begin{axis}[
    tuftelike,
    clip=false,
    axis y line = none,
    xtick = {0, 6},
    xticklabels = {$t_\text{ini}$, $t_\text{ini}+\Delta t$},
    xticklabel style={align=center},
    height=4cm,
    width=.95\textwidth,
    ymin = -0.2,
    ymax = 3.5,
    xmin = 0,
    xmax = 6,
    axis x line shift=35pt,
    ]
    
    \draw[pblue, name path = interpB] plot[domain = 0.0:6.0, variable=\x] (axis cs: {\x}, {3});

    \addplot[mark=x,
    pblue,
    draw=none]
    coordinates{
      (0.1*6,+1+1.5)
      (0.2*6,+0.4+1.5)
      (0.4*6,+1.2+1.5)
      (0.7*6,+1+1.5)
    };

    \addplot[mark=*,
    pblue,
    draw=none]
    coordinates{
      (0,+0.5+1.5) 
      (6,+1.5+1.5)
    }
    node[below left,pos=1,xshift=-1em,yshift=-0.75em]{$c_B$}
    ;

    \draw[porange,domain = 0.0:6.0, name path = interpA, variable=\x] plot (axis cs:{\x}, {1.3});

    \addplot[mark=x,
    porange,
    draw=none]
    coordinates{
      (2,.4) (4,.8) 
    };

    \addplot[mark=*,
    porange,
    draw=none]
    coordinates{
      (0,.5) 
      (6,1.3)
    }
    node[below left,pos=1,xshift=-1em,yshift=-0.5em]{$c_A$}
    ;

    \coordinate(participantA) at (axis cs:0,0);
    \coordinate(participantB) at (axis cs:0,3);

    \coordinate(tsampleA) at (axis cs:0.16*6, 0);  %
    \coordinate(tsampleB_k1) at (axis cs:2, 0);
    \coordinate(tsampleB_k23) at (axis cs:3, 0);
    \coordinate(tsampleB_k4) at (axis cs:4, 0);

    \path[name path= connectionAB] let \p1 = (participantA), \p2 = (participantB), \p3 = (tsampleA) in (\x3,\y1) -- (\x3,\y2);

    \path[name intersections={of=interpA and connectionAB, by=intersectionA}];
    \path[name intersections={of=interpB and connectionAB, by=intersectionBfromA}];

    \path[name path= connectionBA_k1] let \p1 = (participantB), \p2 = (participantA), \p3 = (tsampleB_k1) in (\x3,\y1) -- (\x3,\y2);
    \path[name intersections={of=interpB and connectionBA_k1, by=intersectionB_k1}];
    \path[name intersections={of=interpA and connectionBA_k1, by=intersectionAfromB_k1}];

    \path[name path= connectionBA_k23] let \p1 = (participantB), \p2 = (participantA), \p3 = (tsampleB_k23) in (\x3,\y1) -- (\x3,\y2);
    \path[name intersections={of=interpB and connectionBA_k23, by=intersectionB_k23}];
    \path[name intersections={of=interpA and connectionBA_k23, by=intersectionAfromB_k23}];

    \path[name path= connectionBA_k4] let \p1 = (participantB), \p2 = (participantA), \p3 = (tsampleB_k4) in (\x3,\y1) -- (\x3,\y2);
    \path[name intersections={of=interpB and connectionBA_k4, by=intersectionB_k4}];
    \path[name intersections={of=interpA and connectionBA_k4, by=intersectionAfromB_k4}];

    \node[sample,draw,thick,porange,fill=white, name intersections={of=interpA and connectionAB,total=\t}](sampleA) at (intersectionA){};

    \node[porange,below = 0pt of sampleA]{$c_A\left(t_B^{\alpha}\right)$};

    \draw[->, porange, name intersections={of=interpB and connectionAB,total=\t}](sampleA.north) -- (intersectionBfromA);

    \node[sample,draw,thick,pblue,fill=white, name intersections={of=interpB and connectionBA_k1,total=\t}](sampleB_k1) at (intersectionB_k1){};
    \node[sample,draw,thick,pblue,fill=white, name intersections={of=interpB and connectionBA_k23,total=\t}](sampleB_k23) at (intersectionB_k23){};
    \node[sample,draw,thick,pblue,fill=white, name intersections={of=interpB and connectionBA_k4,total=\t}](sampleB_k4) at (intersectionB_k4){};

    \draw[->, pblue, name intersections={of=interpA and connectionBA_k1,total=\t}](sampleB_k1.south) -- (intersectionAfromB_k1);
    \draw[->, pblue, name intersections={of=interpA and connectionBA_k23,total=\t}](sampleB_k23.south) -- (intersectionAfromB_k23);
    \draw[->, pblue, name intersections={of=interpA and connectionBA_k4,total=\t}](sampleB_k4.south) -- (intersectionAfromB_k4);

    \node[pblue, above = 0pt of sampleB_k1]{$k_1$};
    \node[pblue, above = 0pt of sampleB_k23]{$k_2, k_3$};
    \node[pblue, above = 0pt of sampleB_k4]{$k_4$};

    \draw[dashed] (axis cs: -0.5,1.5) -- node[pos=0,above,pblue]{$B$} node[pos=0,below,porange]{$A$} (axis cs: 6.5,1.5);

    \draw[draw=none] (axis cs:-1,1) rectangle (axis cs: 7, 2);
    
    \end{axis}
    \begin{axis}[
      tuftelike,
      xmin=0,xmax=6,
      clip=false,
      axis y line = none,
      xtick = {0,0.6,1.2,2.4,4.2,6},
      xticklabels = {$t_B^0$,$t_B^1$,$t_B^2$,$t_B^3$,$t_B^4$,$t_B^5$},
      xticklabel style={align=center},
      axis x line*=top,
      xmin = 0,
      xmax = 6,
      height=4cm,
      width=.95\textwidth,
      ]
            \addplot[domain=0:6,draw=none] {1};
    \end{axis}
    \begin{axis}[
      tuftelike,
      xmin=0,xmax=6,
      clip=false,
      axis y line = none,
      xtick = {0,2,4,6},
      xticklabels = {$t_A^0$,$t_A^1$,$t_A^2$,$t_A^3$},
      xticklabel style={align=center},
      axis x line*=bottom,
      xmin = 0,
      xmax = 6,
      height=4cm,
      width=.95\textwidth,
      ]
            \addplot[domain=0:6,draw=none] {1};
    \end{axis}
    \end{tikzpicture}
    \caption{Constant interpolation}
    \label{fig:approaches_a}
  \end{subfigure}
  \begin{subfigure}[b]{0.49\textwidth}
    \centering
        \begin{tikzpicture}
    \begin{axis}[
    tuftelike,
    clip=false,
    axis y line = none,
    xtick = {0, 6},
    xticklabels = {$t_\text{ini}$, $t_\text{ini}+\Delta t$},
    xticklabel style={align=center},
    height=4cm,
    width=.95\textwidth,
    ymin = -0.2,
    ymax = 3.5,
    xmin = 0,
    xmax = 6,
    axis x line shift=35pt,
    ]
    
    \draw[pblue, name path = interpB]
      (axis cs:0,+0.5+1.5) --
      (axis cs:6,+1.5+1.5);

    \addplot[mark=x,
    pblue,
    draw=none]
    coordinates{
      (0.1*6,+1+1.5)
      (0.2*6,+0.4+1.5)
      (0.4*6,+1.2+1.5)
      (0.7*6,+1+1.5)
    };

    \addplot[mark=*,
    pblue,
    draw=none]
    coordinates{
      (0,+0.5+1.5) 
      (6,+1.5+1.5)
    }
    node[below left,pos=1,xshift=-1em,yshift=-0.75em]{$c_B$}
    ;

    \draw[porange, name path = interpA]
      (axis cs:0,.5) --
      (axis cs:6,1.3);

    \addplot[mark=x,
    porange,
    draw=none]
    coordinates{
      (2,.4) (4,.8)
    };

    \addplot[mark=*,
    porange,
    draw=none]
    coordinates{
      (0,.5) 
      (6,1.3)
    }
    node[below left,pos=1,xshift=-1em,yshift=-0.5em]{$c_A$}
    ;

    \coordinate(participantA) at (axis cs:0,0);
    \coordinate(participantB) at (axis cs:0,3);

    \coordinate(tsampleA) at (axis cs:0.16*6, 0);
    \coordinate(tsampleB_k1) at (axis cs:2, 0);
    \coordinate(tsampleB_k23) at (axis cs:3, 0);
    \coordinate(tsampleB_k4) at (axis cs:4, 0);

    \path[name path= connectionAB] let \p1 = (participantA), \p2 = (participantB), \p3 = (tsampleA) in (\x3,\y1) -- (\x3,\y2);

    \path[name intersections={of=interpA and connectionAB, by=intersectionA}];
    \path[name intersections={of=interpB and connectionAB, by=intersectionBfromA}];

    \path[name path= connectionBA_k1] let \p1 = (participantB), \p2 = (participantA), \p3 = (tsampleB_k1) in (\x3,\y1) -- (\x3,\y2);
    \path[name intersections={of=interpB and connectionBA_k1, by=intersectionB_k1}];
    \path[name intersections={of=interpA and connectionBA_k1, by=intersectionAfromB_k1}];

    \path[name path= connectionBA_k23] let \p1 = (participantB), \p2 = (participantA), \p3 = (tsampleB_k23) in (\x3,\y1) -- (\x3,\y2);
    \path[name intersections={of=interpB and connectionBA_k23, by=intersectionB_k23}];
    \path[name intersections={of=interpA and connectionBA_k23, by=intersectionAfromB_k23}];

    \path[name path= connectionBA_k4] let \p1 = (participantB), \p2 = (participantA), \p3 = (tsampleB_k4) in (\x3,\y1) -- (\x3,\y2);
    \path[name intersections={of=interpB and connectionBA_k4, by=intersectionB_k4}];
    \path[name intersections={of=interpA and connectionBA_k4, by=intersectionAfromB_k4}];

    \node[sample,draw,thick,porange,fill=white, name intersections={of=interpA and connectionAB,total=\t}](sampleA) at (intersectionA){};

    \node[porange,below = 0pt of sampleA]{$c_A\left(t_B^{\alpha}\right)$};

    \draw[->, porange, name intersections={of=interpB and connectionAB,total=\t}](sampleA.north) -- (intersectionBfromA);

    \node[sample,draw,thick,pblue,fill=white, name intersections={of=interpB and connectionBA_k1,total=\t}](sampleB_k1) at (intersectionB_k1){};
    \node[sample,draw,thick,pblue,fill=white, name intersections={of=interpB and connectionBA_k23,total=\t}](sampleB_k23) at (intersectionB_k23){};
    \node[sample,draw,thick,pblue,fill=white, name intersections={of=interpB and connectionBA_k4,total=\t}](sampleB_k4) at (intersectionB_k4){};

    \draw[->, pblue, name intersections={of=interpA and connectionBA_k1,total=\t}](sampleB_k1.south) -- (intersectionAfromB_k1);
    \draw[->, pblue, name intersections={of=interpA and connectionBA_k23,total=\t}](sampleB_k23.south) -- (intersectionAfromB_k23);
    \draw[->, pblue, name intersections={of=interpA and connectionBA_k4,total=\t}](sampleB_k4.south) -- (intersectionAfromB_k4);
    
    \node[pblue, above = 0pt of sampleB_k1]{$k_1$};
    \node[pblue, above = 0pt of sampleB_k23]{$k_2, k_3$};
    \node[pblue, above = 0pt of sampleB_k4]{$k_4$};

    \draw[dashed] (axis cs: -0.5,1.5) -- node[pos=0,above,pblue]{$B$} node[pos=0,below,porange]{$A$} (axis cs: 6.5,1.5);

    \draw[draw=none] (axis cs:-1,1) rectangle (axis cs: 7, 2);
    
    \end{axis}
    \begin{axis}[
      tuftelike,
      xmin=0,xmax=6,
      clip=false,
      axis y line = none,
      xtick = {0,0.6,1.2,2.4,4.2,6},
      xticklabels = {$t_B^0$,$t_B^1$,$t_B^2$,$t_B^3$,$t_B^4$,$t_B^5$},
      xticklabel style={align=center},
      axis x line*=top,
      xmin = 0,
      xmax = 6,
      height=4cm,
      width=.95\textwidth,]
            \addplot[domain=0:6,draw=none] {1};
    \end{axis}
    \begin{axis}[
      tuftelike,
      xmin=0,xmax=6,
      clip=false,
      axis y line = none,
      xtick = {0,2,4,6},
      xticklabels = {$t_A^0$,$t_A^1$,$t_A^2$,$t_A^3$},
      xticklabel style={align=center},
      axis x line*=bottom,
      xmin = 0,
      xmax = 6,
      height=4cm,
      width=.95\textwidth,]
            \addplot[domain=0:6,draw=none] {1};
    \end{axis}
    \end{tikzpicture}
    \caption{Linear interpolation}
    \label{fig:approaches_b}
  \end{subfigure}
  \\[1em]
  \begin{subfigure}[b]{0.49\textwidth}
    \centering
        \begin{tikzpicture}
    \begin{axis}[
    tuftelike,
    clip=false,
    axis y line = none,
    xtick = {0, 6},
    xticklabels = {$t_\text{ini}$, $t_\text{ini}+\Delta t$},
    xticklabel style={align=center},
    height=4cm,
    width=.95\textwidth,
    ymin = -0.2,
    ymax = 3.5,
    xmin = 0,
    xmax = 6,
    axis x line shift=35pt,
    ]
    
    \draw[pblue, name path = interpB]
      (axis cs:0,+0.5+1.5) --
      (axis cs:0.1*6,+1+1.5) --
      (axis cs:0.2*6,+0.4+1.5) --
      (axis cs:0.4*6,+1.2+1.5) --
      (axis cs:0.7*6,+1+1.5) --
      (axis cs:6,+1.5+1.5);

    \addplot[mark=*,
    pblue,
    draw=none]
    coordinates{
      (0,+0.5+1.5) 
      (0.1*6,+1+1.5)
      (0.2*6,+0.4+1.5)
      (0.4*6,+1.2+1.5)
      (0.7*6,+1+1.5)
      (6,+1.5+1.5)
    }
    node[below left,pos=1,xshift=-1em,yshift=-0.75em]{$c_B$}
    ;

    \draw[porange, name path = interpA]
      (axis cs:0,.5) --
      (axis cs:2,.4) --
      (axis cs:4,.8) --
      (axis cs:6,1.3);

    \addplot[mark=*,
    porange,
    draw=none]
    coordinates{
      (0,.5) 
      (2,.4)
      (4,.8)
      (6,1.3)
    }
    node[below left,pos=1,xshift=-1em,yshift=-0.5em]{$c_A$}
    ;

    \coordinate(participantA) at (axis cs:0,0);
    \coordinate(participantB) at (axis cs:0,3);

    \coordinate(tsampleA) at (axis cs:0.16*6, 0);
    \coordinate(tsampleB_k1) at (axis cs:2, 0);
    \coordinate(tsampleB_k23) at (axis cs:3, 0);
    \coordinate(tsampleB_k4) at (axis cs:4, 0);

    \path[name path= connectionAB] let \p1 = (participantA), \p2 = (participantB), \p3 = (tsampleA) in (\x3,\y1) -- (\x3,\y2);

    \path[name intersections={of=interpA and connectionAB, by=intersectionA}];
    \path[name intersections={of=interpB and connectionAB, by=intersectionBfromA}];

    \path[name path= connectionBA_k1] let \p1 = (participantB), \p2 = (participantA), \p3 = (tsampleB_k1) in (\x3,\y1) -- (\x3,\y2);
    \path[name intersections={of=interpB and connectionBA_k1, by=intersectionB_k1}];
    \path[name intersections={of=interpA and connectionBA_k1, by=intersectionAfromB_k1}];

    \path[name path= connectionBA_k23] let \p1 = (participantB), \p2 = (participantA), \p3 = (tsampleB_k23) in (\x3,\y1) -- (\x3,\y2);
    \path[name intersections={of=interpB and connectionBA_k23, by=intersectionB_k23}];
    \path[name intersections={of=interpA and connectionBA_k23, by=intersectionAfromB_k23}];

    \path[name path= connectionBA_k4] let \p1 = (participantB), \p2 = (participantA), \p3 = (tsampleB_k4) in (\x3,\y1) -- (\x3,\y2);
    \path[name intersections={of=interpB and connectionBA_k4, by=intersectionB_k4}];
    \path[name intersections={of=interpA and connectionBA_k4, by=intersectionAfromB_k4}];

    \node[sample,draw,thick,porange,fill=white, name intersections={of=interpA and connectionAB,total=\t}](sampleA) at (intersectionA){};

    \node[porange,below = 0pt of sampleA]{$c_A\left(t_B^{\alpha}\right)$};

    \draw[->, porange, name intersections={of=interpB and connectionAB,total=\t}](sampleA.north) -- (intersectionBfromA);

    \node[sample,draw,thick,pblue,fill=white, name intersections={of=interpB and connectionBA_k1,total=\t}](sampleB_k1) at (intersectionB_k1){};
    \node[sample,draw,thick,pblue,fill=white, name intersections={of=interpB and connectionBA_k23,total=\t}](sampleB_k23) at (intersectionB_k23){};
    \node[sample,draw,thick,pblue,fill=white, name intersections={of=interpB and connectionBA_k4,total=\t}](sampleB_k4) at (intersectionB_k4){};

    \draw[->, pblue, name intersections={of=interpA and connectionBA_k1,total=\t}](sampleB_k1.south) -- ([yshift=+3pt]intersectionAfromB_k1);
    \draw[->, pblue, name intersections={of=interpA and connectionBA_k23,total=\t}](sampleB_k23.south) -- ([yshift=+0.5pt]intersectionAfromB_k23);
    \draw[->, pblue, name intersections={of=interpA and connectionBA_k4,total=\t}](sampleB_k4.south) -- ([yshift=+3pt]intersectionAfromB_k4);
    
    \node[pblue, above = 0pt of sampleB_k1]{$k_1$};
    \node[pblue, above = 0pt of sampleB_k23]{$k_2, k_3$};
    \node[pblue, above = 0pt of sampleB_k4]{$k_4$};

    \draw[dashed] (axis cs: -0.5,1.5) -- node[pos=0,above,pblue]{$B$} node[pos=0,below,porange]{$A$} (axis cs: 6.5,1.5);

    \draw[draw=none] (axis cs:-1,1) rectangle (axis cs: 7, 2);
    
    \end{axis}
    \begin{axis}[
      tuftelike,
      xmin=0,xmax=6,
      clip=false,
      axis y line = none,
      xtick = {0,0.6,1.2,2.4,4.2,6},
      xticklabels = {$t_B^0$,$t_B^1$,$t_B^2$,$t_B^3$,$t_B^4$,$t_B^5$},
      xticklabel style={align=center},
      axis x line*=top,
      xmin = 0,
      xmax = 6,
      height=4cm,
      width=.95\textwidth,]
            \addplot[domain=0:6,draw=none] {1};
    \end{axis}
    \begin{axis}[
      tuftelike,
      xmin=0,xmax=6,
      clip=false,
      axis y line = none,
      xtick = {0,2,4,6},
      xticklabels = {$t_A^0$,$t_A^1$,$t_A^2$,$t_A^3$},
      xticklabel style={align=center},
      axis x line*=bottom,
      xmin = 0,
      xmax = 6,
      height=4cm,
      width=.95\textwidth,]
            \addplot[domain=0:6,draw=none] {1};
    \end{axis}
    \end{tikzpicture}
    \caption{Piecewise linear interpolation}
    \label{fig:approaches_c}
  \end{subfigure}
  \begin{subfigure}[b]{0.49\textwidth}
    \centering
        \begin{tikzpicture}
    \begin{axis}[
    tuftelike,
    clip=false,
    axis y line = none,
    xtick = {0, 6},
    xticklabels = {$t_\text{ini}$, $t_\text{ini}+\Delta t$},
    xticklabel style={align=center},
    height=4cm,
    width=.95\textwidth,
    ymin = -0.2,
    ymax = 3.5,
    xmin = 0,
    xmax = 6,
    axis x line shift=35pt,
    ]
    
    \draw[pblue, name path = interpB] plot[domain = 0.0:0.9000000000000001, variable=\x] (axis cs: {\x}, {-1.9727299100667193*\x*\x + 2.016971279373367*\x - 1.5 + 3.5}) -- plot[domain = 0.9000000000000001:1.8000000000000003, variable=\x] (axis cs: {\x}, {1.5868871482447644*\x*\x - 4.3903394255873138*\x + 1.3832898172322916 + 3.5})-- plot[domain = 1.8000000000000003:3.3, variable=\x] (axis cs: {\x}, {-0.59907165651290673*\x*\x + 3.4791122715404565*\x - 5.6992167101827516 + 3.5})-- plot[domain = 3.3:6.0, variable=\x] (axis cs: {\x}, {0.20903845533960438*\x*\x - 1.854414466686092*\x + 3.101102407891057 + 3.5});

    \addplot[mark=*,
    pblue,
    draw=none]
    coordinates{
      (0,+0.5+1.5) 
      (0.1*6,+1+1.5)
      (0.2*6,+0.4+1.5)
      (0.4*6,+1.2+1.5)
      (0.7*6,+1+1.5)
      (6,+1.5+1.5)
    }
    node[below left,pos=1,xshift=-1em,yshift=-0.75em]{$c_B$}
    ;

    \draw[porange,domain = 0.0:6.0, name path = interpA, variable=\x] plot (axis cs:{\x}, {-0.0083333333333333783*\x*\x*\x + 0.11250000000000043*\x*\x - 0.24166666666666747*\x + 0.5});

    \addplot[mark=*,
    porange,
    draw=none]
    coordinates{
      (0,.5) 
      (2,.4)
      (4,.8)
      (6,1.3)
    }
    node[below left,pos=1,xshift=-1em,yshift=-0.5em]{$c_A$}
    ;

    \coordinate(participantA) at (axis cs:0,0);
    \coordinate(participantB) at (axis cs:0,3);

    \coordinate(tsampleA) at (axis cs:0.16*6, 0);
    \coordinate(tsampleB_k1) at (axis cs:2, 0);
    \coordinate(tsampleB_k23) at (axis cs:3, 0);
    \coordinate(tsampleB_k4) at (axis cs:4, 0);

    \path[name path= connectionAB] let \p1 = (participantA), \p2 = (participantB), \p3 = (tsampleA) in (\x3,\y1) -- (\x3,\y2);

    \path[name intersections={of=interpA and connectionAB, by=intersectionA}];
    \path[name intersections={of=interpB and connectionAB, by=intersectionBfromA}];

    \path[name path= connectionBA_k1] let \p1 = (participantB), \p2 = (participantA), \p3 = (tsampleB_k1) in (\x3,\y1) -- (\x3,\y2);
    \path[name intersections={of=interpB and connectionBA_k1, by=intersectionB_k1}];
    \path[name intersections={of=interpA and connectionBA_k1, by=intersectionAfromB_k1}];

    \path[name path= connectionBA_k23] let \p1 = (participantB), \p2 = (participantA), \p3 = (tsampleB_k23) in (\x3,\y1) -- (\x3,\y2);
    \path[name intersections={of=interpB and connectionBA_k23, by=intersectionB_k23}];
    \path[name intersections={of=interpA and connectionBA_k23, by=intersectionAfromB_k23}];

    \path[name path= connectionBA_k4] let \p1 = (participantB), \p2 = (participantA), \p3 = (tsampleB_k4) in (\x3,\y1) -- (\x3,\y2);
    \path[name intersections={of=interpB and connectionBA_k4, by=intersectionB_k4}];
    \path[name intersections={of=interpA and connectionBA_k4, by=intersectionAfromB_k4}];

    \node[sample,draw,thick,porange,fill=white, name intersections={of=interpA and connectionAB,total=\t}](sampleA) at (intersectionA){};

    \node[porange,below = 0pt of sampleA]{$c_A\left(t_B^{\alpha}\right)$};

    \draw[->, porange, name intersections={of=interpB and connectionAB,total=\t}](sampleA.north) -- (intersectionBfromA);

    \node[sample,draw,thick,pblue,fill=white, name intersections={of=interpB and connectionBA_k1,total=\t}](sampleB_k1) at (intersectionB_k1){};
    \node[sample,draw,thick,pblue,fill=white, name intersections={of=interpB and connectionBA_k23,total=\t}](sampleB_k23) at (intersectionB_k23){};
    \node[sample,draw,thick,pblue,fill=white, name intersections={of=interpB and connectionBA_k4,total=\t}](sampleB_k4) at (intersectionB_k4){};

    \draw[->, pblue, name intersections={of=interpA and connectionBA_k1,total=\t}](sampleB_k1.south) -- ([yshift=+3pt]intersectionAfromB_k1);
    \draw[->, pblue, name intersections={of=interpA and connectionBA_k23,total=\t}](sampleB_k23.south) -- ([yshift=+0.5pt]intersectionAfromB_k23);
    \draw[->, pblue, name intersections={of=interpA and connectionBA_k4,total=\t}](sampleB_k4.south) -- ([yshift=+3pt]intersectionAfromB_k4);
    
    \node[pblue, above = 0pt of sampleB_k1]{$k_1$};
    \node[pblue, above = 0pt of sampleB_k23]{$k_2, k_3$};
    \node[pblue, above = 0pt of sampleB_k4]{$k_4$};

    \draw[dashed] (axis cs: -0.5,1.5) -- node[pos=0,above,pblue]{$B$} node[pos=0,below,porange]{$A$} (axis cs: 6.5,1.5);

    \draw[draw=none] (axis cs:-1,1) rectangle (axis cs: 7, 2);
    
    \end{axis}
    \begin{axis}[
      tuftelike,
      xmin=0,xmax=6,
      clip=false,
      axis y line = none,
      xtick = {0,0.6,1.2,2.4,4.2,6},
      xticklabels = {$t_B^0$,$t_B^1$,$t_B^2$,$t_B^3$,$t_B^4$,$t_B^5$},
      xticklabel style={align=center},
      axis x line*=top,
      xmin = 0,
      xmax = 6,
      height=4cm,
      width=.95\textwidth,]
            \addplot[domain=0:6,draw=none] {1};
    \end{axis}
    \begin{axis}[
      tuftelike,
      xmin=0,xmax=6,
      clip=false,
      axis y line = none,
      xtick = {0,2,4,6},
      xticklabels = {$t_A^0$,$t_A^1$,$t_A^2$,$t_A^3$},
      xticklabel style={align=center},
      axis x line*=bottom,
      xmin = 0,
      xmax = 6,
      height=4cm,
      width=.95\textwidth,]
            \addplot[domain=0:6,draw=none] {1};
    \end{axis}
    \end{tikzpicture}
    \caption{B-spline interpolation}
    \label{fig:approaches_d}
  \end{subfigure}
}
\caption{Comparison of different multi-rate coupling strategies. Two participants $A$ and $B$ use three and five time steps in the time window $[t_{\text{ini}}, t_{\text{ini}} + \Delta t]$, respectively. $A$ uses constant time steps and the forth-order Runge-Kutta RK4 method. $B$ use adaptive time steps and the generalized-$\alpha$ method. Interpolants are computed from data values at different points in time and sampled at example locations to retrieve data values for the coupling partner. Notation is formally introduced in \Cref{sec:wi}.}
\label{fig:approaches}
\end{figure}
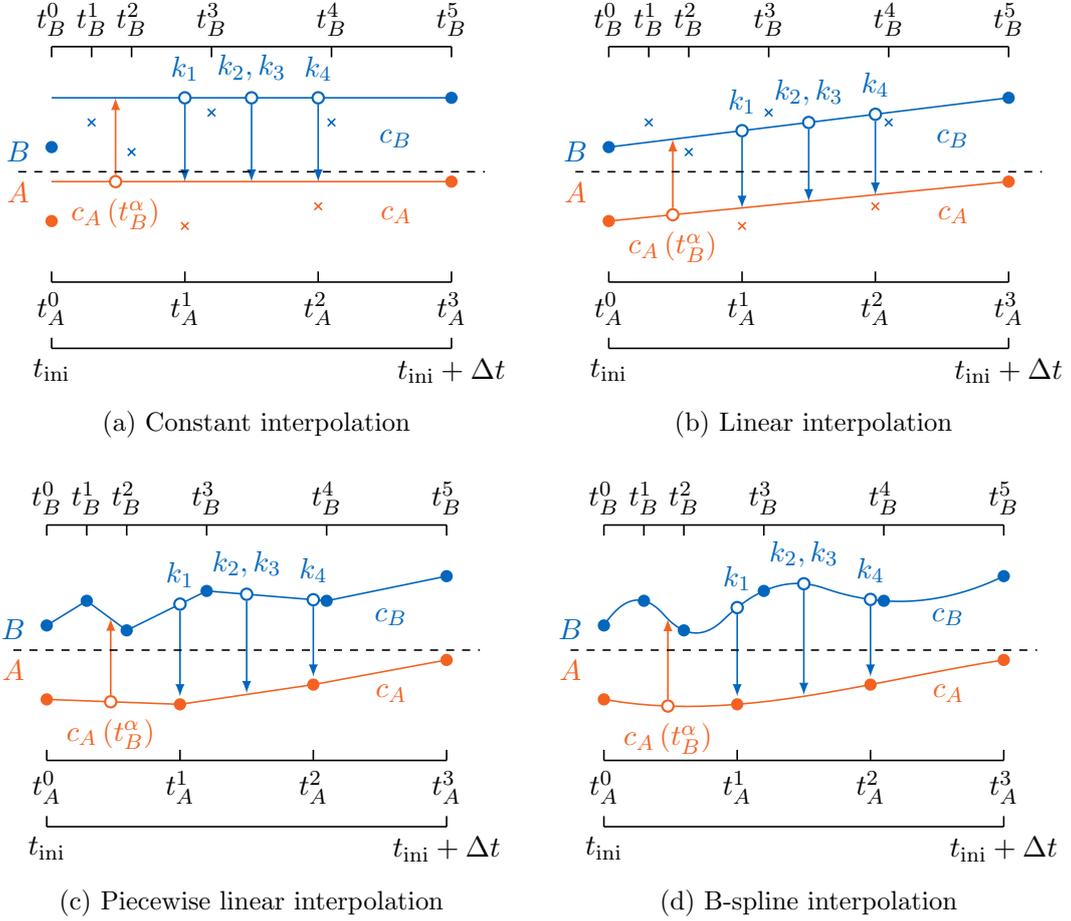

For the sake of stability and simplicity, many black-box coupling strategies adopt the smallest time step across all components and synchronize after each step, e.g. \cite{Nonino2022, Breuer2012}. While straightforward, this approach wastes computational effort on slower-evolving processes. An alternative is to synchronize only after a larger, fixed time window, allowing each component to progress at its own rate. Such a multi-rate setup is the focus of this paper. \Cref{fig:approaches} illustrates various multi-rate coupling strategies using two components with three and five time steps per window. Each coupling window can either be executed once (or a fixed number of times) or repeated until a convergence criterion is met---referred to as explicit and implicit coupling, respectively. Additionally, components may advance sequentially or concurrently, denoted as serial and parallel coupling.

The most basic multi-rate approach (\Cref{fig:approaches_a}) involves exchanging only the most recent coupling values. However, this na\"ive method often suffers from significant coupling errors and stability issues. It also typically reduces the convergence order of the time integrators, often to only linear convergence. A simple improvement involves linear interpolation without increasing data exchange (\Cref{fig:approaches_b}), as discussed by De Moerloose et al.~in the context of FSI \cite{DeMoerloose2019}. Other simple alternatives exist. Strang splitting \cite{Strang2007}, for example, allows coupling two components, whose time step sizes differ by a factor of two. If the individual time steps are executed in the right order, quadratic convergence order can be achieved. Again in the context of FSI, variants of this approach are discussed by Farhat and Lesoinne \cite{Farhat1998}. Nonetheless, such splitting methods are limited in handling arbitrary multi-rate configurations.

Waveform iteration is a promising general-purpose black-box technique, which uses arbitrary interpolants, so-called waveforms, within each window. For instance, \Cref{fig:approaches_c,fig:approaches_d} depict the usage of piecewise linear and third-degree B-spline interpolation, respectively. With appropriate interpolants, it is possible to preserve the original convergence order of the time integrators \cite{Matthies2003}. Originally developed for parallelizing integrated circuit simulations in electrical engineering \cite{Gander2015}, waveform iteration is now applied in co-simulation and PDE contexts (e.g., see \cite{Garcia2021,Tang2021,Monge2019a}). In prior work, we have combined waveform iteration with quasi-Newton acceleration methods for FSI \cite{Rueth2021}.

Nevertheless, applying waveform iteration generically in a black-box fashion to PDE-based coupled simulations is nontrivial. Supporting general multi-rate configurations —including adaptive time stepping and multiple components—requires sophisticated communication logic. Furthermore, the method must integrate smoothly with existing building blocks for coupled PDE problems, such as data mapping for non-matching coupling meshes and fixed-point acceleration for strongly-coupled problems. To address these challenges, we explore the integration of waveform iteration into the black-box coupling library preCICE in this paper, summarizing key parts of the dissertation of Rodenberg \cite{Rodenberg2025}.

preCICE \cite{Chourdakis2022} follows a peer-to-peer library approach for coupling. This means that the coupled components---referred to as participants in preCICE---call the coupler as a library and directly communicate with each other, not requiring any central orchestrator. The library provides various data mapping techniques, ranging from simple projections to advanced partition-of-unity radial-basis-function interpolation \cite{Schneider2025}, as well as multiple fixed-point acceleration methods, from basic underrelaxation to sophisticated quasi-Newton schemes \cite{Mehl2016}. Communication between participants is asynchronous and parallel, using MPI or TCP/IP. The library includes ready-to-use adapters for several simulation codes (e.g., OpenFOAM \cite{Chourdakis2023}, FEniCS \cite{Rodenberg2021}) and offers APIs in multiple languages, such as C++, Fortran, and Python. Its high-level API hides the underlying communication logic, exposing it through a single call to advance the coupling at each time step. While other coupling libraries exist---such as OpenPALM \cite{Duchaine2015}, MUI \cite{Tang2015}, YAC \cite{Hanke2016}, or DTK \cite{Slattery2016}---they typically leave the coupling logic, fixed-point acceleration, and time interpolation to the user.

In the remainder of this paper, we describe the mathematical framework of waveform iteration in \Cref{sec:wi}. Afterwards, \Cref{sec:implementation} describes the implementation in preCICE in detail, covering the only minimal necessary changes to the user interface, the data layout, and the interplay with fixed-point acceleration, data mapping, and communication. \Cref{sec:results} then presents numerical results showcasing that the new coupling approach can handle complex multi-rate setups and can achieve higher-order integration in time, significantly reducing numerical errors and, thus, enhancing preCICE, benefiting its extensive user community.

\section{Waveform iteration}
\label{sec:wi}

We consider two coupled participants, labeled $A$ and $B$. While the setup is formulated for a pair of coupled participants, the implementation is designed to support an arbitrary number. Each participant operates on its respective spatial domain, denoted $ \Omega_A $ and $ \Omega_B $. These domains may be identical, partially overlapping, or only in contact at a shared interface. The region of interaction between the two participants is referred to as the coupling domain $ \Omega_A \cap \Omega_B $.
Typically, $A$ and $B$ use different meshes to discretize the coupling domain. Data mappings are required to translate coupling data between these meshes. For simplicity, we omit these mapping here (cf. \cite{Schneider2025, Chourdakis2022} for further discussion of data mapping in preCICE).

The temporal domain is divided into a sequence of time windows. In the following, we focus on a single such window, given by $[t_{\text{ini}}, t_{\text{ini}} + \Delta t]$, where $ t_{\text{ini}} $ is the initial time and $ \Delta t $ the size of the window.
Within this time window, each participant performs time integration using its own scheme, which may be explicit or implicit and may feature adaptive time stepping. If needed, we denote any time step size by $\delta_t$. The only requirement is that both participants synchronize at the final time of the window. The time steps taken by participants $A$ and $B$ are denoted
\[
t_A^0, t_A^1, \ldots, t_A^{n_A} \quad \text{and} \quad t_B^0, t_B^1, \ldots, t_B^{n_B},
\]
where $ t_A^0 = t_B^0 = t_{\text{ini}} $ and $ t_A^{n_A} = t_B^{n_B} = t_{\text{ini}} + \Delta t $.
The number of time steps $n_A$ and $n_B$ might be different in different time windows.
At the end of each of their time steps, participants output coupling data:
\[
c_A^1, c_A^2, \ldots, c_A^{n_A} \in \mathbb{R}^{d_A}, \quad
c_B^1, c_B^2, \ldots, c_B^{n_B} \in \mathbb{R}^{d_B},
\]
which are high-dimensional vectors encoding all coupling-relevant quantities.

To obtain continuous representations of the coupling data over the entire time window, we apply interpolation to these time-discrete outputs. This yields the functions:
\[
c_A: [t_{\text{ini}}, t_{\text{ini}} + \Delta t] \to \mathbb{R}^{d_A}, \quad
c_B: [t_{\text{ini}}, t_{\text{ini}} + \Delta t] \to \mathbb{R}^{d_B}.
\]
Both interpolants can be evaluated by the respective participants at arbitrary time points, not just the discrete steps, allowing for use at intermediate stages such as those arising in Runge-Kutta methods.

We define the participant operators
\[
\mathcal{A}: c_B \mapsto (c_A^1, \ldots, c_A^{n_A}), \quad
\mathcal{B}: c_A \mapsto (c_B^1, \ldots, c_B^{n_B}),
\]
which represent the discrete-time output of one participant in response to continuous-time input from the other. Additionally, we define interpolation operators (we omit the subscript when not referring to a specific side):
\[
\mathcal{I}_{c^0}: (c^1, \ldots, c^n) \mapsto c,
\]
which reconstruct a continuous function $ c $ given an initial value $ c^0 $ and a set of discrete-time samples.
Combining the participant and interpolation operators yields the time-continuous operators
\[
\widehat{\mathcal{A}} = \mathcal{I}_{c_A^0} \circ \mathcal{A}, \quad
\widehat{\mathcal{B}} = \mathcal{I}_{c_B^0} \circ \mathcal{B},
\]
which map continuous inputs to continuous outputs.

In the case of serial coupling and if participant $B$ advances first, the coupling procedure can be expressed as a time-continous fixed-point equation:
\[
c_A = \widehat{\mathcal{A}} \circ \widehat{\mathcal{B}}\,(c_A).
\]
Alternatively, the same relationship can be written in terms of discrete-time outputs:
\[
(c_A^1, \ldots, c_A^{n_A}) = \mathcal{A} \circ \mathcal{I}_{c_B^0} \circ \mathcal{B} \circ \mathcal{I}_{c_A^0}(c_A^1, \ldots, c_A^{n_A}).
\]

To ensure stability and accuracy, an implicit coupling strategy is employed by iterating on this fixed-point equation. Denoting the iteration index by $k$, the basic fixed-point iteration becomes:
\[
[(c_A^1, \ldots, c_A^{n_A})]^{k+1} = \mathcal{A} \circ \mathcal{I}_{c_B^0} \circ \mathcal{B} \circ \mathcal{I}_{c_A^0}([(c_A^1, \ldots, c_A^{n_A})]^k).
\]

In practice, simple fixed-point iteration is often insufficient for convergence \cite{Rueth2018}. To enhance performance, we introduce an acceleration operator $\mathcal{Q}$, leading to the iteration:
\[
[(c_A^1, \ldots, c_A^{n_A})]^{k+1} = \mathcal{Q} \circ \mathcal{A} \circ \mathcal{I}_{c_B^0} \circ \mathcal{B} \circ \mathcal{I}_{c_A^0}([(c_A^1, \ldots, c_A^{n_A})]^k).
\]
In its simplest form, $ \mathcal{Q} $ may represent an under-relaxation operator, though more sophisticated variants such as quasi-Newton acceleration can also be employed (cf. \cite{Rueth2021}).

We summarize the serial-implicit coupling algorithm (with $B$ advancing first) as follows. Tildes (e.g., $ \tilde{c}_A^1 $) denote values prior to acceleration:\\

\begin{enumerate}[itemsep=0.22cm]
  \item Initialize from previous window, obtain $c_A^0$ and $c_B^0$ and set constant initial guess $\left[c_A\right]^0 \equiv c_A^0$, $k=0$.
  \item Advance $B$ (with $n_B$ time steps):
	      $\left[\left(c_B^1, \ldots, c_B^{n_B}\right)\right]^{k+1} = \mathcal{B} \left(\left[c_A\right]^k\right)$.
  \item Interpolate: $\left[c_B\right]^{k+1} = \mathcal{I}_{c_B^0} \left(\left[\left(c_B^1, \ldots, c_B^{n_B}\right)\right]^{k+1} \right)$.
  \item Advance $A$ (with $n_A$ time steps):
	      $\left[\left(\tilde{c}_A^1, \ldots, \tilde{c}_A^{n_A}\right)\right]^{k+1} = \mathcal{A} \left(\left[c_B\right]^{k+1}\right)$.
	\item If $\left\| \left[\tilde{c}_A^{n_A}\right]^{k+1} - \left[c_A^{n_A}\right]^{k}\right\|$ small enough: Go to next window.
	\item Accelerate: $\left[\left(c_A^1, \ldots, c_A^{n_A}\right)\right]^{k+1} = \mathcal{Q} \left(\left[\left(\tilde{c}_A^1, \ldots, \tilde{c}_A^{n_A}\right)\right]^{k+1}\right)$.
  \item Interpolate: $\left[c_A\right]^{k+1} = \mathcal{I}_{c_A^0} \left(\left[\left(c_A^1, \ldots, c_A^{n_A}\right)\right]^{k+1} \right)$.
  \item Increment $k$ and go back to step 2.
\end{enumerate}
\vspace{0.4cm}

For the interpolation, several options are possible. To keep the user interface simple, we do not make use of time derivatives. As a general and powerful choice, we employ B-spline interpolation \cite{Dierckx1993} of degree $p$. For implementation simplicity, we restrict the interpolation to the current time window, which imposes $p \le n$. If insufficient data points are available, the interpolation degree is automatically reduced. A possible workaround is to let the participant produce dense output, if possible, and sub-sample additional points \cite{Rodenberg2025}. Both the restriction on the current time window and the avoidance of time derivatives could be relaxed in future work.

\section{Implementation}
\label{sec:implementation}

This section presents the main contribution of our work: the implementation of waveform iteration in the coupling library preCICE. We first outline the minimal changes to the user interface in \Cref{ssec:ui}. \Cref{ssec:data-layout} describes the internal data layout, and \Cref{ssec:data-flow} focuses on the data flow within preCICE, highlighting the interaction with essential building blocks such as communication, data mapping, and fixed-point acceleration. While our examples again focus on two participants, $A$ and $B$, the implementation supports arbitrary numbers of participants.

\subsection{User interface}
\label{ssec:ui}

A central design philosophy of preCICE is to provide a high-level API, where a single call to advance progresses the coupling for a time step. This single call encapsulates data mapping, fixed-point acceleration, and communication, abstracting away the low-level send and receive operations found in other coupling libraries. This enables users to configure the coupling scheme, including who sends data to whom, at runtime.

\begin{figure}[h!]
  \centering
    \includegraphics[width=0.75\textwidth]{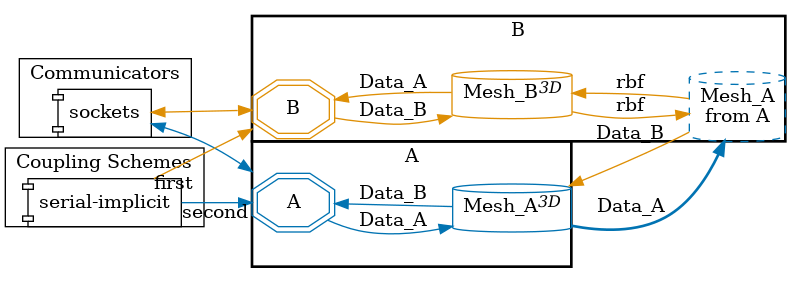}
  \caption{Visualization of preCICE configuration example of \Cref{code:config} using the preCICE CLI
  }
  \label{fig:config}
\end{figure}

\begin{listing}[h!]
  \caption{Minimal preCICE example in Python}
  \label{code:example}
\small
\begin{minted}[mathescape,linenos,numbersep=5pt,gobble=0,frame=none,framesep=20mm,escapeinside=||,breaklines]{python}
import precice

participant = precice.Participant("A", "../precice-config.xml", ...)

vertices = extract_coupling_mesh() # shape "number vertices $\times$ mesh dimension"
vertex_ids = participant.set_mesh_vertices("Mesh_A", vertices)

if participant.requires_initial_data:
    data_A = initialize_data() # shape "number vertices $\times$ data dimension"
    participant.write_data("Mesh_A", "Data_A", vertex_ids, data_A)

participant.initialize()

while participant.is_coupling_ongoing():
    if participant.requires_writing_checkpoint():
        write_checkpoint()

    solver_dt = compute_adaptive_dt()
    precice_dt = participant.get_max_time_step_size()
    dt = min(solver_dt, precice_dt)

    data_B = participant.read_data("Mesh_A", "Data_B", vertex_ids, dt)
    data_A = solve_A(data_B, dt)
    participant.write_data("Mesh_A", "Data_A", vertex_ids, data_A)
    participant.advance(dt)

    if participant.requires_reading_checkpoint():
        read_checkpoint()
\end{minted}
\end{listing}

preCICE is written in C++, but we use the Python bindings in the following to explain the user-facing interface. A minimal example from the perspective of participant $A$ is shown in \Cref{code:example}. The \pyinline{Participant} object is created in Line 3 using a configuration file, given in \Cref{code:config} and visualized in \Cref{fig:config}. Coupling meshes are defined next; users can define arbitrary numbers of meshes. The data initialization step is optional and described in more detail in \Cref{ssec:data-flow}. Calling \pyinline{initialize} exchanges meshes between participants $A$ and $B$, prepares data mappings, and establishes communication channels.

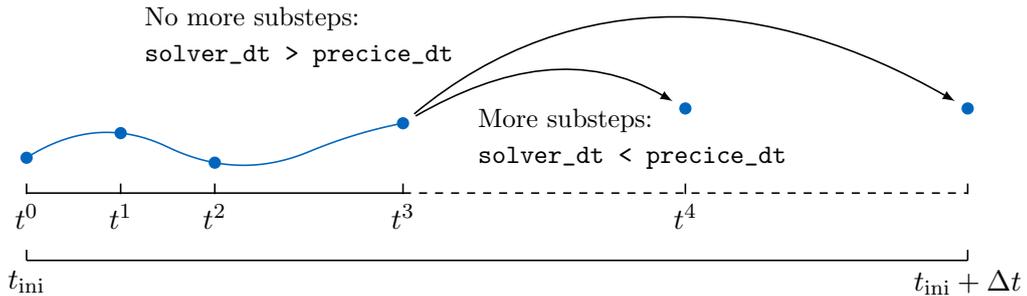
\begin{figure}[h!]
  \center
      \begin{tikzpicture}
    \begin{axis}[
    tuftelike,
    clip=false,
    axis y line = none,
    xtick = {0, 6},
    xticklabels = {$t_\text{ini}$, $t_\text{ini}+\Delta t$},
    xticklabel style={align=center},
    height=4cm,
    width=.95\textwidth,
    ymin = -0.2,
    ymax = 3.5,
    xmin = 0,
    xmax = 6,
    axis x line shift=35pt,
    ]
    
    \draw[pblue, name path = interpB] plot[domain = 0.0:0.9000000000000001, variable=\x] (axis cs: {\x}, {-1.9727299100667193*\x*\x + 2.016971279373367*\x - 1.5 + 1.5}) -- plot[domain = 0.9000000000000001:1.8000000000000003, variable=\x] (axis cs: {\x}, {1.5868871482447644*\x*\x - 4.3903394255873138*\x + 1.3832898172322916 + 1.5})-- plot[domain = 1.8000000000000003:2.4, variable=\x] (axis cs: {\x}, {-0.59907165651290673*\x*\x + 3.4791122715404565*\x - 5.6992167101827516 + 1.5});

    \addplot[mark=*,
    pblue,
    draw=none]
    coordinates{
      (0,0) 
      (0.1*6,+1-0.5)
      (0.2*6,+0.4-0.5)
      (0.4*6,+1.2-0.5)
      (0.7*6,+1.5-0.5)
      (6,+1.5-0.5)
    };

    \draw[draw=none] (axis cs:-0.5,1) rectangle (axis cs: 6.5, 2);

    \node[circle,draw=none](ssFrom) at (axis cs:0.4*6,+1.2-0.5){};
    \node[circle,draw=none](ssTo1) at (axis cs:0.7*6,+1.5-0.5){};
    \node[circle,draw=none](ssTo2) at (axis cs:6,+1.5-0.5){};

    \draw (ssFrom) 
      edge[->,out=30,in=150] 
      node[right, align=left, anchor=north west,pos=0.2,yshift=-0.1cm]{\small{More substeps:}\\  \small{\pyinline{solver_dt < precice_dt}}}
      (ssTo1);
    \draw (ssFrom) 
      edge[->,out=40,in=150] 
      node[left, align=left, anchor=south east,pos=0.1,xshift=0cm]{\small{No more substeps:} \\ \small{\pyinline{solver_dt > precice_dt}}}
      (ssTo2);

    \end{axis}
    \begin{axis}[
      tuftelike,
      clip=false,
      axis y line = none,
      axis line style={draw=none},
      xtick = {0,0.6,1.2,2.4,4.2,6},
      xticklabels = {$t^0$,$t^1$,$t^2$,$t^3$,$t^4$,},
      xticklabel style={align=center},
      axis x line*=bottom,
      xmin = 0,
      xmax = 6,
      height=4cm,
      width=.95\textwidth,]
      \addplot[domain=0:6,draw=none] {1};

      \draw[black,yshift=-1pt](axis cs:0,0.75) -- (axis cs:2.4,0.75);
      \draw[black,yshift=-1pt, dashed](axis cs:2.4,0.75) -- (axis cs:6,0.75);
      
    \end{axis}
    \end{tikzpicture}
  \caption{Time stepping within a time window $[t_{\text{ini}}, t_{\text{ini}} + \Delta t]$.
  \pyinline{dt = min(solver_dt, precice_dt)} reaches the next time step or the end of the time window.
  }
  \label{fig:time-step-API}
\end{figure}

Each iteration of the while-loop (lines 14 to 28) corresponds to one time step. If implicit coupling is used, the simulation may jump back in time to the start of the time window via checkpointing (lines 15-16 and 27-29). The variable \pyinline{solver_dt} is the time step size desired by the solver, while \pyinline{precice_dt} represents the maximum permissible time step within the current window. The actual time step \pyinline{dt} is chosen as the minimum of these two. See \Cref{fig:time-step-API} for a sketch.

Data is read from and written to internal buffers in lines 22 and 24. The function \pyinline{solve_A} advances the participant $A$ by one time step \pyinline{dt}. The function \pyinline{advance} then progresses the coupling in preCICE, including mapping, acceleration, and communication. For more details, we refer to the official documentation\footnote{\url{https://precice.org/docs.html}}.

\begin{listing}[p]
  \caption{preCICE configuration example. The mesh of $A$, \xmlinline{"Mesh_A"}, is sent from participant $A$ to participant $B$ and data mappings in both directions are computed on participant $B$. The B-spline (waveform) degree $p$ is set to 3 for both data fields. The exchange of substep data, necessary for waveform iteration, is switched on (default for implicit coupling).}
  \label{code:config}
\small
\begin{minted}[mathescape,linenos,numbersep=5pt,gobble=0,frame=none,framesep=20mm,escapeinside=||,breaklines]{xml}
<precice-configuration>

  <data:scalar name="Data_A" waveform-degree="3" />
  <data:scalar name="Data_B" waveform-degree="3" />

  <mesh name="Mesh_A"  dimensions="3">
    <use-data name="Data_A" />
    <use-data name="Data_B" />
  </mesh>

  <mesh name="Mesh_B"  dimensions="3">
    <use-data name="Data_A" />
    <use-data name="Data_B" />
  </mesh>

  <participant name="A">
    <provide-mesh name="Mesh_A" />
    <write-data name="Data_A" mesh="Mesh_A" />
    <read-data name="Data_B" mesh="Mesh_A" />
  </participant>

  <participant name="B">
    <receive-mesh name="Mesh_A" from="A" />
    <provide-mesh name="Mesh_B" />
    <mapping:rbf direction="write" from="Mesh_B" to="Mesh_A" constraint="conservative" />
    <mapping:rbf direction="read" from="Mesh_A" to="Mesh_B" constraint="consistent" />
    <write-data name="Data_B" mesh="Mesh_B" />
    <read-data name="Data_A" mesh="Mesh_B" />
  </participant>

  <m2n:sockets acceptor="A" connector="B" />

  <coupling-scheme:serial-implicit>
    <participants first="B" second="A" />
    <max-time-windows value="2" />
    <time-window-size value="1.0" />
    <max-iterations value="2" />
    <exchange data="Data_A" mesh="Mesh_A" from="A" to="B" substeps="true" initialize="true" />
    <exchange data="Data_B" mesh="Mesh_A" from="B" to="A" substeps="true" />
    <relative-convergence-measure limit="1e-4" data="Data_A" mesh="Mesh_A" />
    <acceleration:IQN-ILS>
      <data name="Data_A" mesh="Mesh_A" />
    </acceleration:IQN-ILS>
  </coupling-scheme:serial-implicit>
</precice-configuration>
\end{minted}
\end{listing}

Prior to waveform iteration, up to version 2 of preCICE, constant interpolation was used as multi-rate strategy (depicted in \Cref{fig:approaches_a}). All data written in a window was overwritten and only the last value retained. \pyinline{read_data} always returned this final value. The time argument used in \Cref{code:example} line 22 did not yet exist.
We gradually ported all major components to support waveform iteration (depicted in \Cref{fig:approaches_c,fig:approaches_d}), an effort we concluded in version 3.2. Now, the time argument in \pyinline{read_data} can be used to specify a relative time within each time step, ranging from \pyinline{0} (corresponding to the time step start) to \pyinline{dt} (the time step end). This allows evaluation at intermediate times, for example at \pyinline{0}, \pyinline{dt/2}, and \pyinline{dt} for the stages of the classical Runge-Kutta scheme.
Data is no longer overwritten but stored and used to construct interpolants. The \xmlinline{waveform-degree} attribute specifies the degree of interpolation---in \Cref{code:config}, cubic B-splines are used---subject to the limitations discussed in \Cref{sec:wi}. If the attribute \xmlinline{substeps="off"} is added to an \xmlinline{exchange} tag, then substep handling and associated communication are turned off, corresponding to \Cref{fig:approaches_b}. While this mode might also have an efficiency sweet spot, it is mainly retained for comparison and may be removed in the future.

\subsection{Data layout}
\label{ssec:data-layout}

The core challenge in implementing waveform iteration is that adaptive time stepping introduces variable numbers of data entries within each time window. In an earlier prototype~\cite{Rueth2021}, we used fixed time step sizes.

\begin{figure}[h!]
  \center
  \begin{tikzpicture}
  \node(mesh){\cppinline{Mesh}};
  \node[right = of mesh](data){\cppinline{Data}};
  \node[right = of data](storage){\cppinline{Storage}};
  \node[right = of storage](stample){\cppinline{Stample}};
  \node[right = of stample](sample){\cppinline{Sample}};
  \node[below right = of stample,yshift=+3em](timestamp){\cppinline{double} $t_A^i$};
  \draw[-](mesh) -- node[pos=0,above right]{\cppinline{1}} node[pos=1,above left]{\cppinline{n}} (data);
  \draw[-](data) -- node[pos=0,above right]{\cppinline{1}} node[pos=1,above left]{\cppinline{1}} (storage);
  \draw[-](storage) -- node[pos=0,above right]{\cppinline{1}} node[pos=1,above left]{\cppinline{m}} (stample);
  \draw[-](stample.east) -- node[pos=0,above right]{\cppinline{1}} node[pos=1,above left]{\cppinline{1}} (sample.west);
  \draw[-](stample.east) -- node[pos=1,below left]{\cppinline{1}} (timestamp.west);

  \node[above=1.5em of mesh, yshift=+0.25em]{\small\xmlinline{"Mesh_A"}};

  \node[above=1.5em of data, yshift=+0.25em]{\small\xmlinline{"Data_A"}};
  \node[above=1.5em of data, yshift=-0.75em]{\small\xmlinline{"Data_B"}};

  \node[stample, above=1.5em of storage, yshift=-0.5em](s1){};
  \node[stample, above=1.5em of storage](s2){};
  \node[stample, above=1.5em of storage, yshift=+0.5em](s3){};

  \begin{pgfonlayer}{bg}
    \node[fill=gray!25, fit=(s1) (s3), rounded corners=5](storage){};
  \end{pgfonlayer}

  \node[stample, above=1.5em of stample]{};
  \node[sample, above=1.5em of sample]{};
  \end{tikzpicture}
  \caption{Core data structures in preCICE and their relationships. The symbols for \cppinline{Storage}, \cppinline{Stample}, and \cppinline{Sample} are used throughout the paper.}
  \label{fig:data-layout}
\end{figure}
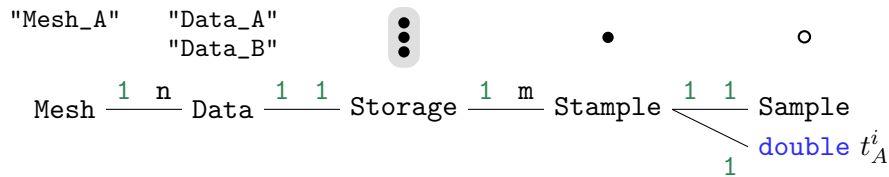

\Cref{fig:data-layout} illustrates the relationship between the core data structures. The class \xmlinline{Mesh} (such as \xmlinline{"Mesh_A"}) can hold multiple \xmlinline{Data} fields (such as \xmlinline{"Data_A"} or \xmlinline{"Data_B"}), as seen in the configuration. Each data field uses a \cppinline{Storage} backend that maintains a time-ordered list of \cppinline{Stample} objects---short for time-stamped samples.

For clarity, we only consider scalar data in the following. In the Python API, data is passed as NumPy arrays; in the C++ backend, this becomes a generic span and is ultimately wrapped into an \cppinline{Eigen::VectorXd} \cite{Eigen2010}, forming the core of a \cppinline{Sample}. Each Stample then holds a \cppinline{Sample} and an associated time stamp. The time stamp is not known when \pyinline{write_data} is called, but only during \pyinline{advance}. Therefore, the implementation splits writing into two steps: first, we collect multiple \cppinline{Sample} objects, then during \pyinline{advance}, we turn each into a \cppinline{Stample}.
This design simplifies the user interface by avoiding the need to manually specify time stamps. The design specifically assumes that written data corresponds to the end of each time step, which matches most time integration schemes. For others, users may need to perform interpolation themselves.
Absolute time stamps are used internally, which simplifies synchronization across participants and potentially multiple coupling schemes with different time windows each.

\begin{figure}[h!]
  \center
  \begin{tabular}{cccc}
    \texttt{setSampleAtTime($t^3$,\begin{tikzpicture}\node[sample](newsample){};\end{tikzpicture})}
    & 
    \texttt{sample($t$)}
    &
    \texttt{trimAfter($t^0$)}
    &
    \texttt{trimBefore($t^2$)} 
    \\
    \begin{tikzpicture}[scale=1.0]

    \coordinate(storage) at (0,0);
    \node[stample, yshift=+1em,label={[label distance=0.5em]left:\small$t^0$}](s1) at (storage){};
    \node[stample,label={[label distance=0.5em]left:\small$t^1$}](s2) at (storage){};
    \node[stample, yshift=-1em,label={[label distance=0.5em]left:\small$t^2$}](s3) at (storage){};
    \node[sample, yshift=-2.25em](s4) at (storage){};

    \begin{pgfonlayer}{bg}
    \node[fill=gray!25, fit=(s1) (s3), rounded corners=5](storageMark){};
    \end{pgfonlayer}

    \coordinate(storage_) at (1,0);

    \node[stample, yshift=+1em,label={[label distance=0.5em]right:\small$t^0$}](s1_) at (storage_){};
    \node[stample,label={[label distance=0.5em]right:\small$t^1$}](s2_) at (storage_){};
    \node[stample, yshift=-1em,label={[label distance=0.5em]right:\small$t^2$}](s3_) at (storage_){};
    \node[stample, yshift=-2em,label={[label distance=0.5em]right:\small$t^3$}](s4_) at (storage_){};

    \begin{pgfonlayer}{bg}
    \node[fill=gray!25, fit=(s1_) (s4_), rounded corners=5](storageMark_){};
    \end{pgfonlayer}

    \draw[->](s2-|storageMark.east) -- (s2_-|storageMark_.west);

    \node[draw=none,circle,inner sep=.35em](s4From)at(s4){};
    \node[draw=none,circle,inner sep=.55em](s2_To)at(s2_){};
    \draw[->](s4From) -- (s2_To);

    \draw[draw=none] (0,0.8) rectangle (1,-1);
    \end{tikzpicture}
&
\begin{tikzpicture}[scale=1.0]

\coordinate(storage) at (0,0);
    \node[stample, yshift=+1em,label={[label distance=0.5em]left:\small$t^0$}](s1) at (storage){};
    \node[stample,label={[label distance=0.5em]left:\small$t^1$}](s2) at (storage){};
    \node[stample, yshift=-1em,label={[label distance=0.5em]left:\small$t^2$}](s3) at (storage){};

\begin{pgfonlayer}{bg}
\node[fill=gray!25, fit=(s1) (s3), rounded corners=5](storageMark){};
\end{pgfonlayer}

\coordinate(sample_) at (1,0);

\node[sample,label={right:\small$t$}](s_) at (sample_){};

    \begin{pgfonlayer}{bg}
    \node[fill=white, fit=(s_), rounded corners=5](sampleMark){};
    \end{pgfonlayer}

\draw[->](storageMark.east) -- (sampleMark.west);

    \draw[draw=none] (0,0.8) rectangle (1,-1);

\end{tikzpicture}
        & 
    \begin{tikzpicture}[scale=1.0]

    \coordinate(storage) at (0,0);
    \node[stample, yshift=+1em,label={[label distance=0.5em]left:\small$t^0$}](s1) at (storage){};
    \node[stample,label={[label distance=0.5em]left:\small$t^1$}](s2) at (storage){};
    \node[stample, yshift=-1em,label={[label distance=0.5em]left:\small$t^2$}](s3) at (storage){};

    \begin{pgfonlayer}{bg}
    \node[fill=gray!25, fit=(s1) (s3), rounded corners=5](storageMark){};
    \end{pgfonlayer}

    \coordinate(storage_) at (1,0);

    \node[stample, yshift=+1em,label={[label distance=0.5em]right:\small$t^0$}](s1_) at (storage_){};

    \begin{pgfonlayer}{bg}
    \node[fill=gray!25, fit=(s1_), rounded corners=5](storageMark_){};
    \end{pgfonlayer}

    \draw[->](s1-|storageMark.east) -- (s1_-|storageMark_.west);

    \draw[draw=none] (0,0.8) rectangle (1,-1);

    \end{tikzpicture}  
        & 
    \begin{tikzpicture}[scale=1.0]

    \coordinate(storage) at (0,0);
    \node[stample, yshift=+1em,label={[label distance=0.5em]left:\small$t^0$}](s1) at (storage){};
    \node[stample,label={[label distance=0.5em]left:\small$t^1$}](s2) at (storage){};
    \node[stample, yshift=-1em,label={[label distance=0.5em]left:\small$t^2$}](s3) at (storage){};

    \begin{pgfonlayer}{bg}
    \node[fill=gray!25, fit=(s1) (s3), rounded corners=5](storageMark){};
    \end{pgfonlayer}

    \coordinate(storage_) at (1,0);

    \node[stample, yshift=-1em,label={[label distance=0.5em]right:\small$t^2$}](s3_) at (storage_){};

    \begin{pgfonlayer}{bg}
    \node[fill=gray!25, fit=(s3_), rounded corners=5](storageMark_){};
    \end{pgfonlayer}

    \draw[->](s3-|storageMark.east) -- (s3_-|storageMark_.west);

    \draw[draw=none] (0,0.8) rectangle (1,-1);

    \end{tikzpicture}

\end{tabular}
  \caption{Methods to add and retrieve data from the \cppinline{Storage} class}
  \label{fig:storage-api}
\end{figure}
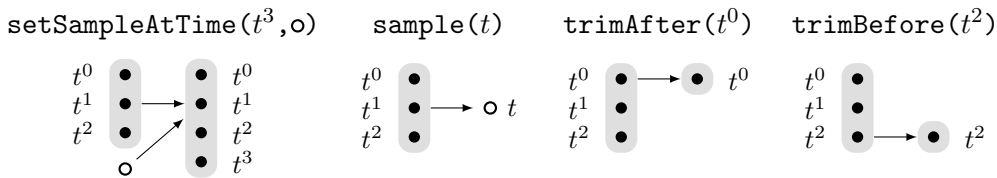

The \cppinline{Storage} collects multiple \cppinline{Stample} objects and provides methods to add and retrieve data, see \Cref{fig:storage-api}. When \cppinline{advance} is called, the new \cppinline{Sample} is converted and stored (using \cppinline{setSampleAtTime}). If a time window converges, all data before the window end is trimmed (using \cppinline{trimBefore}), with the final \cppinline{Stample} retained as new starting data. If the window does not converge, all data after the window start is discarded (using \cppinline{trimAfter}), preserving only the initial \cppinline{Stample}.
When calling \pyinline{read_data}, the relative time is translated to an absolute one and used to sample from the \cppinline{Storage}. To this end, the \cppinline{Storage} holds a B-spline interpolation class, adapted from Eigen's implementation for cache efficiency\footnote{\url{https://github.com/precice/precice/pull/1765\#pullrequestreview-1632933653}}.

\subsection{Data flow}
\label{ssec:data-flow}

While waveform iteration is well known in co-simulation, it is not widely used in black-box coupling of PDE-based solvers. The additional complexity arises because PDE simulations require the integration of multiple building blocks: data mapping, parallel communication, and fixed-point acceleration. We now describe how preCICE manages their interaction under waveform iteration.

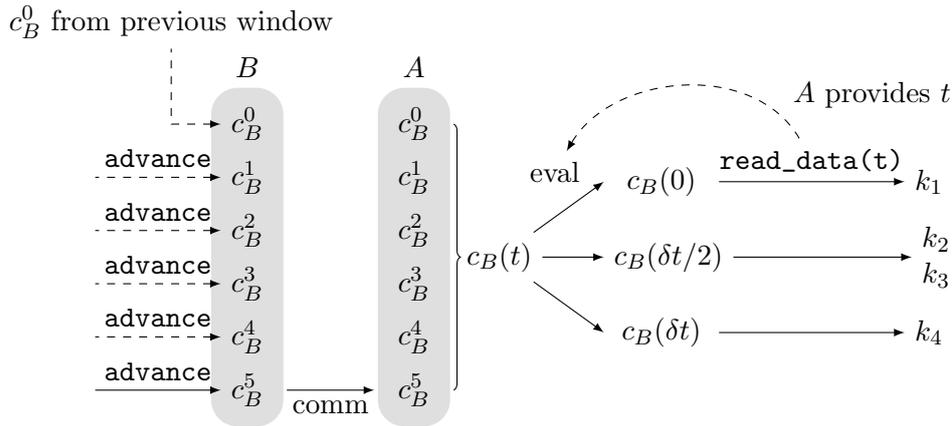
\begin{figure}
  \begin{tikzpicture}

    \node(x0){$c_B^{0}$};
    \node[below = 0cm of x0](x1){$c_B^{1}$};
    \node[below = 0cm of x1](x2){$c_B^{2}$};
    \node[below = 0cm of x2](x3){$c_B^{3}$};
    \node[below = 0cm of x3](x4){$c_B^{4}$};
    \node[below = 0cm of x4](x5){$c_B^{5}$};

    \draw[<-, dashed](x0)-|++(-1,+1) node[above,align=center]{$c^0_B$ from previous window};
    \draw[<-, dashed](x1)--node[above]{\pyinline{advance}} ++ (-2,0);
    \draw[<-, dashed](x2)--node[above]{\pyinline{advance}} ++(-2,0);
    \draw[<-, dashed](x3)--node[above]{\pyinline{advance}} ++(-2,0);
    \draw[<-, dashed](x4)--node[above]{\pyinline{advance}} ++(-2,0);
    \draw[<-](x5)--node[above,align=center]{\pyinline{advance}} ++(-2,0);

    \node[right = 1.5cm of x0](xcomm0){$c_B^{0}$};
    \node[right = 1.5cm of x1](xcomm1){$c_B^{1}$};
    \node[right = 1.5cm of x2](xcomm2){$c_B^{2}$};
    \node[right = 1.5cm of x3](xcomm3){$c_B^{3}$};
    \node[right = 1.5cm of x4](xcomm4){$c_B^{4}$};
    \node[right = 1.5cm of x5](xcomm5){$c_B^{5}$};

    \draw[decoration={brace,raise=5pt},decorate]
  (xcomm0.east) -- node[right=6pt](xcommt) {$c_B(t)$} (xcomm5.east);

  \coordinate(xmid) at ($0.5*(xcomm0) + 0.5*(xcomm5)$);

  \node[right = 2.5cm of xmid,yshift=1cm, minimum width=1.5cm, align=center](xcommtk1){$c_B(0)$};
  \node[right = 2.5cm of xmid, minimum width=1.5cm, align=center](xcommtk23){$c_B(\delta t / 2)$};
  \node[right = 2.5cm of xmid,yshift=-1cm, minimum width=1.5cm, align=center](xcommtk4){$c_B(\delta t)$};

  \draw[->](xcommt) -- node[pos=0.8,above left](evallabel){eval} (xcommtk1.west);
  \draw[->](xcommt) -- (xcommtk23.west);
  \draw[->](xcommt) -- (xcommtk4.west);

  \draw[->](xcommtk1) -- node[above](readlabel){\pyinline{read_data(t)}} ++(3.2,0) node[right,align=left]{$k_1$};
  \draw[->](xcommtk23) -- ++(3.2,0) node[right,align=left]{$k_2$\\$k_3$};
  \draw[->](xcommtk4) -- ++(3.2,0) node[right,align=left]{$k_4$};
  
  \draw[->, dashed](readlabel) [out=120,in=60]to node[above right,pos=0.1] (tlabel){$A$ provides $t$} (evallabel);

  \begin{pgfonlayer}{bg}
    \node[fill=gray!25, dotted,fit=(x0) (x5), rounded corners=10,label={above:$B$}](storage){};
    \node[fill=gray!25, dotted,fit=(xcomm0) (xcomm5), rounded corners=10,label={above:$A$}](storagecomm){};
    \coordinate(topbox)at($(storage.north)+(0,1.5em)$);
  \end{pgfonlayer}

  \draw[->]([xshift=.1em]x5-|storage.east) -- node[below](commlabel){comm} ([xshift=-.1em]xcomm5-|storagecomm.west);

\end{tikzpicture}
  \caption{Fundamental design decision of the implementation. Communicatin of storage from $B$ to $A$ at the end of a time window (here, after 5 time steps) and construction of waveform $c_B$ on reading participant $A$.
  }
  \label{fig:read-interpolation}
\end{figure}

A fundamental design decision is to determine which participant creates the waveform---the writing or the reading participant. This choice affects all components. For instance, if participant $B$ writes \xmlinline{"Data_B"} and participant $A$ reads it, then in the first option we communicate the entire Storage to $A$, who then constructs the waveform locally. In the second option, $B$ constructs the waveform and sends only the evaluated samples to $A$. The latter would require $A$ to communicate sampling times to $B$, either in \pyinline{read_data} or in the previous \pyinline{advance}. This either contradicts the preCICE design of only communicating in \pyinline{advance} or impairs usability with adaptive time stepping. Therefore, we choose the first option: waveform construction occurs on the reader side as depicted in \Cref{fig:read-interpolation}. This does increase communication volume when $B$ uses more time steps than $A$ requires, but the opposite direction might cause more performance issues. For mitigation, we want to study data compression in future work.

For data mapping, preCICE separates the typically costly initialization based on the meshes (in initialize) and the actual data-dependend mapping in advance. The latter is performed at the window end. Similarly as in previous preCICE versions \cite{Chourdakis2022}, the write mapping (e.g., \xmlinline{"Data_B"} from \xmlinline{"Mesh_B"} to \xmlinline{"Mesh_A"} on B) is applied before communication; the read mapping (e.g., \xmlinline{"Data_A"} from \xmlinline{"Mesh_A"} to \xmlinline{"Mesh_B"} on B) is applied afterward. With waveform iteration now, each \cppinline{Sample} is individually mapped, which does, however, not increase the initialization cost. The sample at the window start is excluded, as it was already mapped before---either in the previous time window or the previous iteration.

Communication in preCICE builds on repartitioning the coupling meshes to establish parallel communication channels \cite{Uekermann2016}. This step as well as the actual communication is independent of waveform iteration, only more data is now communicated. Storage objects are serialized; due to dynamic size, the message size must be communicated beforehand. For simplicity, we also transmit the initial sample in each window, though this could be optimized.

Coupling schemes define the order in which coupled participants advance. Both serial (one after the other) and parallel (both simultaneously) coupling are supported. The extension to waveform iteration is conceptually straightforward, with full storages now communicated. A slight complication is initialization. If \xmlinline{initialize="true"} (as in \Cref{code:config} for \xmlinline{"Data_A"}), initial values are exchanged. In parallel coupling, both sides can initialize and exchange data immediately. In serial coupling, only the second participant ($A$) previously needed the final value from the first ($B$) to start. Now, the initial value is also needed to construct a first waveform, so $B$ must write it, and $A$ receives it normally as part of the first \cppinline{Storage} communication in the first advance. Only if \xmlinline{substeps="off"}, then an extra communication step is needed. We omit technical details here, but refer to Rodenberg~\cite{Rodenberg2025}.

Acceleration is the most complex component to adapt. preCICE supports several variants, from simple underrelaxation to advanced quasi-Newton methods. They all rely on tracking histories of iterates and residuals, which becomes problematic under adaptive time stepping, as time grids differ. In our earlier prototype \cite{Rueth2021}, we therefore considered fixed time grids only.
Two main solutions are possible: always interpolating all history data to the current time grid or using a fixed auxiliary grid. The latter is far less computationally expensive. Still, the additional interpolation leads to additional computational cost and introduces an additional numerical error, which can be controlled by refining the auxiliary grid.
This approach was implemented and analyzed independently by Kotarsky and Birken \cite{Kotarsky2025}, also showing the generality and modularity of our implementation. Their evaluation compared different time grids and concluded that reusing the grid from the first iterate suffices. This also supports the fixed grid case automatically. They further demonstrate the efficiency benefits of adaptive stepping, which is why we do not present such results here. We refer the reader to their work for further details.

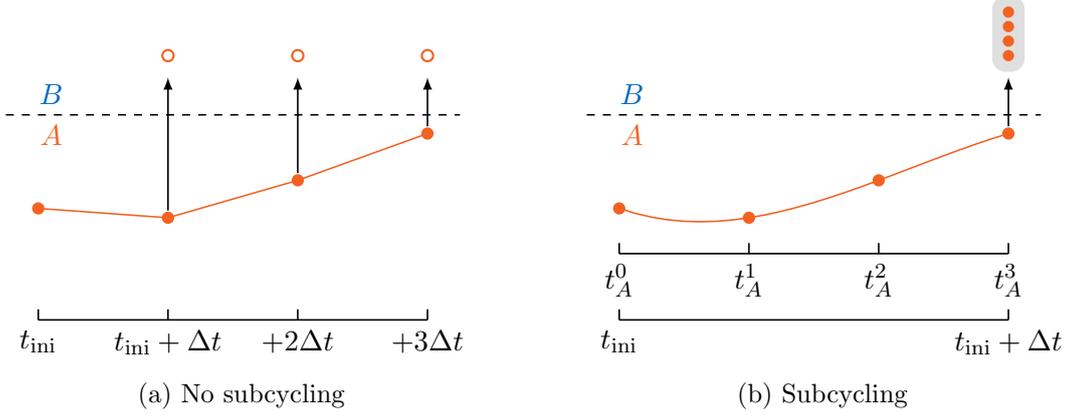
\begin{figure}[h!]
  \begin{subfigure}[t]{.48\textwidth}
    \begin{tikzpicture}
    \begin{axis}[
    tuftelike,
    clip=false,
    width=.95\textwidth,
    height=.4\textwidth,
    axis y line = none,
    xtick = {0, 2, 4, 6},
    xticklabels = {$t_\text{ini}$, $t_\text{ini}+\Delta t$, $+2\Delta t$,$+3\Delta t$},
    ymin = 0.3,
    ymax = 1.3,
    axis x line shift=35pt,
    ]

    \addplot[tension=1,
    mark=*,
    porange,
    draw]
    coordinates{
      (0,.5)
      (2,.4)
      (4,.8)
      (6,1.3)
    };

    \draw[dashed] (axis cs: -0.5,1.5) -- node[pos=0.1,above,pblue]{$B$} node[pos=0.1,below,porange]{$A$} (axis cs: 6.5,1.5);

    \draw[<-] (axis cs: 2,1.9) -- coordinate[yshift=0.75em,pos=0](s1) ([yshift=0.25em]axis cs: 2,0.4);
    \draw[<-] (axis cs: 4,1.9) -- coordinate[yshift=0.75em,pos=0](s2) ([yshift=0.25em]axis cs: 4,0.8);
    \draw[<-] (axis cs: 6,1.9) -- coordinate[yshift=0.75em,pos=0](s3) ([yshift=0.25em]axis cs: 6,1.3);

    \node[sample, draw=porange] at (s1.north) {};
    \node[sample, draw=porange] at (s2.north) {};
    \node[sample, draw=porange] at (s3.north) {};

    \draw[draw=none] (axis cs:-1,1) rectangle (axis cs: 7, 2);

    \end{axis}
    \end{tikzpicture}
    \caption{No subcycling}
    \label{sfig:commnosubcycling}
    \end{subfigure}
    \hfill
    \begin{subfigure}[t]{.48\textwidth}
    \begin{tikzpicture}
    \begin{axis}[
    tuftelike,
    clip=false,
    width=.95\textwidth,
    height=.4\textwidth,
    axis y line = none,
    xtick = {0, 6},
    xticklabels = {$t_\text{ini}$, $t_\text{ini}+\Delta t$},
    xticklabel style={align=center},
    ymin = 0.3,
    ymax = 1.3,
    axis x line shift=35pt,
    ]

    \draw[porange,domain = 0.0:6.0, variable=\x] plot (axis cs:{\x}, {-0.0083333333333333783*\x*\x*\x + 0.11250000000000043*\x*\x - 0.24166666666666747*\x + 0.5});

    \addplot[mark=*,
    porange,
    draw=none]
    coordinates{
      (0,.5)
      (2,.4)
      (4,.8)
      (6,1.3)
    };

    \draw[dashed] (axis cs: -0.5,1.5) -- node[pos=0.1,above,pblue]{$B$} node[pos=0.1,below,porange]{$A$} (axis cs: 6.5,1.5);

    \draw[<-] (axis cs: 6,1.9) -- coordinate[yshift=0.75em,pos=0](storagepos) ([yshift=0.25em]axis cs: 6,1.3);

    \node[stample, porange] (s3) at (storagepos) {};
    \node[stample, porange] (s2) at ([yshift=+0.5em]storagepos) {};
    \node[stample, porange] (s1) at ([yshift=+1em]storagepos) {};
    \node[stample, porange] (s0) at ([yshift=+1.5em]storagepos) {};

    \begin{pgfonlayer}{bg}
      \node[fill=gray!25, fit=(s0) (s3), rounded corners=5](storage){};
    \end{pgfonlayer}

    \draw[draw=none] (axis cs:-1,1) rectangle (axis cs: 7, 2);

    \end{axis}
    \begin{axis}[
      tuftelike,
      xmin=0,xmax=6,
      clip=false,
      axis y line = none,
      xtick = {0,2,4,6},
      xticklabels = {$t_A^0$,$t_A^1$,$t_A^2$,$t_A^3$},
      xticklabel style={align=center},
      axis x line*=bottom,
      ymin = 0.3,
      ymax = 1.3,
      xmin = 0,
      xmax = 6,
      width=.95\textwidth,
      height=.4\textwidth,
      ]
            \addplot[domain=0:6,draw=none] {1};
    \end{axis}
    \end{tikzpicture}
    \caption{Subcycling}
    \label{sfig:commsubcycling}
    \end{subfigure}
  \caption{Illustration of data used in communication and acceleration with $A$ using identical time step and window size $\Delta t$ (\Cref{sfig:commnosubcycling}) and $A$ using time steps samller than window size $\Delta t$ (\Cref{sfig:commsubcycling}). In \Cref{sfig:commnosubcycling}, $A$ sends, maps, and accelerates a single \cppinline{Sample} at the end of each time window. In \Cref{sfig:commsubcycling}, $A$ performs $3$ time steps and sends, maps, and accelerates the resulting \cppinline{Storage}.}
  \label{fig:comm}
\end{figure}

To conclude the section, we compare the computational cost of two scenarios with identical time step sizes, which therefore require no interpolation: (i) one where the time window equals the time step size ($\Delta t = \delta t$, \Cref{sfig:commnosubcycling}), and (ii) one with subcycling ($\Delta t = n \cdot \delta t$, \Cref{sfig:commsubcycling}). Assuming both variants converge in $k$ iterations, each requires $k \cdot n$ mapping operations. Communication costs are $k \cdot n$ in the first case and $k \cdot (n+1)$ in the second, although this overhead could be reduced. Subcycling, however, decreases the number of synchronization points, which is particularly beneficial when individual time steps have differing runtimes.
How this reduced synchronization influences the number of iterations is investigated in \Cref{sec:results}.
For acceleration, the cost depends on the chosen method. All acceleration techniques scale linearly with the number of degrees of freedom (e.g., \cite{Scheufele2017}). In the case of quasi-Newton acceleration combined with waveform iteration, we distinguish between a full variant, which uses all substep data to compute the update, and a reduced variant, which relies only on data at the end of each time window \cite{Kotarsky2025,Rueth2021}. While the full variant is expected to incur comparable costs, the reduced variant is significantly cheaper.

\section{Results}
\label{sec:results}

We analyze three academic test cases, all taken from the preCICE tutorials\footnote{\url{https://precice.org/tutorials.html}}. Although simple, these cases demonstrate the significant advantages of waveform iteration over earlier multi-rate approaches implemented in preCICE. Detailed setups and results are provided in a separate data publication \cite{Rodenberg2025widata}.

\subsection{Partitioned oscillator}

\begin{figure}[h!]
  \begin{subfigure}{.48\textwidth}
      \begin{tikzpicture}
\tikzstyle{spring}=[thick,decorate,decoration={zigzag,pre length=0.3cm,post length=0.3cm,segment length=6}]
\tikzstyle{ground}=[fill,pattern=north east lines,draw=none,minimum width=0.3cm,minimum height=0.2cm]
\tikzstyle{mass}=[draw, fill=gray!25, thick, minimum width=0.7cm, minimum height=0.7cm]
\node (W1) [ground, rotate=-90, minimum width=2cm] {};
\node (M1) [mass, fill=porange!50, right= 1.3cm of W1.north] {$m_A$};
\draw [dashed] (M1.north) -- ++ (0,.5);
\draw [->] ($(M1.north) + (0,.25)$) -- node[above]{$u_A$} ++ (1,0);
\draw [spring] (W1.north) -- node[below]{$k_A$}(M1.west);
\node (M2) [mass, fill=pblue!50, right= 1.3cm of M1] {$m_B$};
\draw [dashed] (M2.north) -- ++ (0,.5);
\draw [->] ($(M2.north) + (0,.25)$) -- node[above]{$u_B$} ++ (1,0);
\draw [spring] (M1.east) -- node[below]{$k_{AB}$}(M2.west);
\node (W2) [ground, rotate=90, minimum width=2cm, right= 1.3cm of M2, xshift=-1cm] {};
\draw [spring] (M2.east) -- node[below]{$k_B$}(W2.north);
\end{tikzpicture}
      \vspace{1em}
      \caption{Schematic drawing}
      \label{sfig:oscillator_schematic}
  \end{subfigure}
  \hfill
  \begin{subfigure}{.48\textwidth}
      \centering
      \begin{tikzpicture}%
      \begin{axis}[
      tuftelike,
      domain=0:1,
      samples=100,
      legend style={at={(.5,1.2)},anchor=north, draw=none},
      legend cell align=left,
      width= \textwidth,
      height= .5\textwidth,
      ymin=-1,
      ]

      \addplot[no marks] {0.5*(cos(deg(2*pi*x))+cos(deg(6*pi*x)))};
      \addlegendentry{$u_A$}
      \addplot[no marks, dashed] {0.5*(cos(deg(2*pi*x))-cos(deg(6*pi*x)))};
      \addlegendentry{$u_B$}

      \end{axis}
      \end{tikzpicture}
      \caption{Displacement $u_A$ and $u_B$ over time $t$}
      \label{sfig:oscillator_analytical}
    \end{subfigure}
    \vspace{-0.35cm}
    \caption{Oscillating two-mass-spring-system (left) and its analytical solution (right)}
    \label{fig:oscillator}
\end{figure}
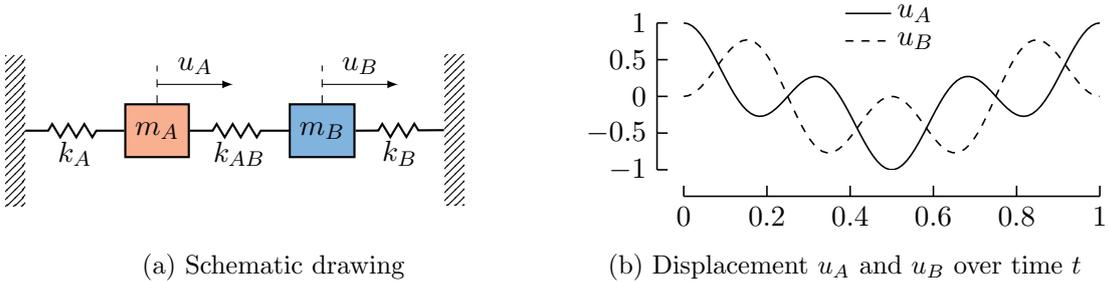

Schüller et al.~\cite{Schueller2022} introduce a simple partitioned oscillator model, which we use to illustrate the core ideas of waveform iteration, validate our implementation, and investigate fundamental behavior. The system is modeled by two coupled ODEs describing the dynamics of a dimensionless two-mass-spring setup, as shown in \Cref{sfig:oscillator_schematic}:
\begin{align}
\label{eq:osciA}
m_A \ddot{u}_A + \left(k_A + k_{AB}\right) u_A - k_{AB} u_B &= 0\;,\\
\label{eq:osciB}
m_B \ddot{u}_B - k_{AB} u_A + \left(k_B + k_{AB}\right) u_B &= 0\;.
\end{align}
We use $k_A = k_B = 4 \pi^2,\, k_{AB} = 16 \pi^2$, $m_A = m_B = 1$, and the initial conditions
\begin{align*}
u_A(t=0)&=1\;,& \dot{u}_A(t=0)&=0\;,&u_B(t=0)&=0\;,&\dot{u}_B(t=0)&=0\;.
\end{align*}
\Cref{sfig:oscillator_analytical} shows the analytical solution for this specific case (cf.\ \cite{Schueller2022}).

\begin{figure}[h!]
  \centering
  \begin{tikzpicture}
    \tikzstyle{spring}=[thick,decorate,decoration={zigzag,pre length=0.3cm,post length=0.3cm,segment length=6}]
    \tikzstyle{ground}=[fill,pattern=north east lines,draw=none,minimum width=0.3cm,minimum height=0.2cm]
    \tikzstyle{mass}=[draw, fill=gray!25, thick, minimum width=0.7cm, minimum height=0.7cm]
    \node (W1) [ground, rotate=-90, minimum width=2cm] {};
    \node (M1) [mass, fill=pblue!50, right= 1.3cm of W1.north] {$m_A$};
    \node (M2fake) [circle, draw=black, fill=porange!50, right= 1.3cm of M1] {};

    \draw [dashed] (M2fake.north) -- ++ (0,.5);
    \draw [->] ($(M2fake.north) + (0,.25)$) -- node[above](U2ToM1){$u_B$} ++ (1,0);
    \draw [dashed] (M1.north) -- ++ (0,.5);
    \draw [->] ($(M1.north) + (0,.25)$) -- node[above](U1FromM1){$u_A$} ++ (1,0);
    \draw [spring] (W1.north) -- node[below]{$k_A$}(M1.west);
    \draw [spring] (M1.east) -- node[below]{$k_{AB}$}(M2fake.west);

    \node (M1fake) [circle, draw=black, fill=pblue!50, right= 1cm of M2fake] {};
    \node (M2) [mass, fill=porange!50, right= 1.3cm of M1fake] {$m_B$};
    \draw [dashed] (M2.north) -- ++ (0,.5);
    \draw [->] ($(M2.north) + (0,.25)$) -- node[above](U2FromM2){$u_B$} ++ (1,0);
    \draw [dashed] (M1fake.north) -- ++ (0,.5);
    \draw [->] ($(M1fake.north) + (0,.25)$) -- node[above](U1ToM2){$u_A$} ++ (1,0);

    \node (W2) [ground, rotate=90, minimum width=2cm, right= 1.3cm of M2, xshift=-1cm] {};
    \draw [spring] (M2.east) -- node[below]{$k_B$}(W2.north);
    \draw [spring] (M1fake.east) -- node[below]{$k_{AB}$}(M2.west);

    \draw[](U2FromM2.north) edge[->,out=135, in=45] node[above right]{ghost layer for $m_A$} (U2ToM1.north);
    \draw[](U1FromM1.north) edge[->,out=45, in=135] node[above left]{ghost layer for $m_B$} (U1ToM2.north);

\end{tikzpicture}
  \vspace{1em}
  \caption{Partitioning of the two-mass-spring system}
  \label{fig:oscillator_schwarz}
\end{figure}
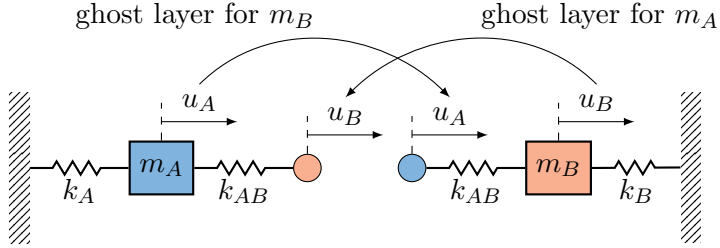

To study coupling strategies, we artificially partition the system following an overlapping Schwarz approach, illustrated in \Cref{fig:oscillator_schwarz}. This means we solve \Cref{eq:osciA,eq:osciB} independently using separate ODE solvers, each relying on approximations for $u_B$ and $u_A$, respectively. We examine the interval $t\in\left[0,1\right]$, covering one full oscillation period, for various time window and time step sizes. For both displacements $u_A$ and $u_B$, we compute the maximum error across all time steps, denoted by $e_A$ and $e_B$.

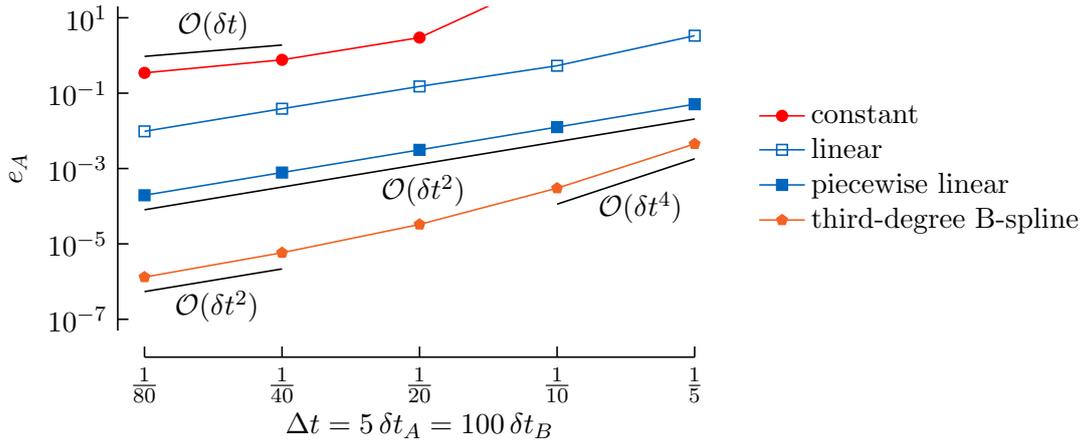
\begin{figure}
  \center
  \begin{tikzpicture}[
 mark options={solid}
 ]
\begin{loglogaxis}[
 height=.4\textwidth,
 width=.6\textwidth,
 tuftelike,
 legend cell align=left,
 xlabel={$\Delta t = 5 \, \delta t_A = 100 \, \delta t_B$},
 x label style={at={(axis description cs:0.5,-0.1)},anchor=north},
 ylabel={$e_A$},
 y label style={at={(axis description cs:-0.025,0.5)},anchor=south},
 align=center,
 legend style={at={(1.1,.5)},anchor=west, draw=none, align=left},
 xmin=0.0125,
 xmax=0.2,
 ymax=20,
 ymin=5e-8,
 xtick = {0.2, 0.1, 0.05, 0.025, 0.0125},
 xticklabels = {$\frac{1}{5}$,$\frac{1}{10}$, $\frac{1}{20}$, $\frac{1}{40}$, $\frac{1}{80}$},
 ]
    \addplot[mark=*, color=red] table[x=time window size, y=error Mass-Left, col sep=comma]{figures/OscillatorVariants/data/constant.csv};
    \addlegendentry{constant}

    \addplot[mark=square,color=pblue] table[x=time window size , y=error Mass-Left, col sep=comma]{figures/OscillatorVariants/data/linear.csv};
    \addlegendentry{linear}

    \addplot[mark=square*,color=pblue] table[x=time window size , y=error Mass-Left, col sep=comma]{figures/OscillatorVariants/data/piecewise_linear.csv};
    \addlegendentry{piecewise linear}

    \addplot[mark=pentagon*,color=porange] table[x=time window size , y=error Mass-Left, col sep=comma]{figures/OscillatorVariants/data/third_degree_b-spline.csv};
    \addlegendentry{third-degree B-spline}

    \draw[] ([yshift=+0.75em]axis cs:0.025,0.5) -- node[above]{$\mathcal{O}(\delta t)$} ([yshift=+0.75em]axis cs:0.0125,0.25);
    \draw[] ([yshift=-0.5em]axis cs:0.2,0.0499905460271616) -- node[below,xshift=+.05cm]{$\mathcal{O}(\delta t^2)$} ([yshift=-0.5em]axis cs:0.0125,0.0001952755704186);
    \draw[] ([yshift=-0.5em]axis cs:0.025,5.257798539659574e-06) -- node[below,xshift=+.05cm]{$\mathcal{O}(\delta t^2)$} ([yshift=-0.5em]axis cs:0.0125,1.3144496349148935e-06);
    \draw[] ([yshift=-0.5em]axis cs:0.2,0.0044) -- node[below,xshift=+.5em]{$\mathcal{O}(\delta t^4)$} ([yshift=-0.5em]axis cs:0.1,0.000275);

    \end{loglogaxis}
    \end{tikzpicture}
  \caption{Illustrative comparison of the four multi-rate approaches introduced in \Cref{fig:approaches} for the partitioned oscillator. $A$ uses RK4 with $\delta t_A = \Delta t / 5$, $B$ uses GA with $\delta t_B = \Delta t / 100$.}
  \label{fig:multirate_oscillator_compare_variants}
\end{figure}

In a first experiment, we compare the four multi-rate coupling approaches introduced in \Cref{fig:approaches}. As in that figure, participant $A$ employs the RK4 method, while participant $B$ uses the generalized Alpha (GA) method. $A$ and $B$ use $5$ and $100$ constant time steps per time window, respectively; this extreme ratio is chosen to clearly highlight the impact of the method. \Cref{fig:multirate_oscillator_compare_variants} shows the resulting errors for decreasing time window sizes. In all oscillator experiments, we use serial-implicit coupling and iterate until the relative coupling error (defined as the maximum change in $u_A$ and $u_B$ between successive iterations) drops below $10^{-10}$. No acceleration method is needed to stabilize the coupling.

As expected, subcycling with constant interpolation yields first-order convergence, and with linear interpolation, second-order convergence. Waveform iteration with piecewise-linear interpolation also achieves second-order convergence, but with considerably smaller errors. When using third-degree B-spline interpolation, waveform iteration exhibits fourth-order convergence for large time windows, where RK4 in $A$ dominates, and second-order convergence for smaller windows, where GA in $B$ dominates.

\begin{figure}[h!]
  \centering
  \pgfmathdeclarefunction{lg10}{1}{%
    \pgfmathparse{ln(#1)/ln(10)}%
}

\begin{tikzpicture}[mark options={solid}, use Hobby shortcut]
    \begin{loglogaxis}[
    tuftelike,
    xlabel={$m_B$ uses {GA} with $\delta t_B$},
    x label style={at={(axis description cs:0.5,-0.1)},anchor=north},
    ylabel={$m_A$ uses {RK4} with $\delta t_A$},
    y label style={at={(axis description cs:-0.025,0.5)},anchor=south},
    width=.8\textwidth,
    xmax=2e-2,
    xmin=1e-5,
    xtick={1e-5,1e-4,1e-3,1e-2,1e-1},
    xticklabels={$10^{-5}$, $10^{-4}$,$10^{-3}$, $10^{-2}$,},
    ymin=1e-3,
    ymax=3e-2,
    ytick={1e-3,1e-2,1e-1},
    yticklabels={$10^{-3}$, $10^{-2}$,},
    axis equal image,
    colormap name=viridis,
    point meta max =-3,
    point meta min =-8,
    colorbar,
    colorbar style={
        title={$\max(e_A,e_B)$},
        ymax = -3,
        ymin = -8,
        ytick={-3,-4,-5,-6,-7,-8},
        yticklabels={$10^{-3}$,$10^{-4}$,$10^{-5}$,$10^{-6}$,$10^{-7}$,$10^{-8}$},
    }
    ]

    \addplot[scatter,only marks, scatter src=explicit] table[x=x, y=y, meta=z, meta expr=lg10(\thisrow{z}), col sep=semicolon]{figures/ErrorContours/script/cdata/times_reduced.csv};

    \begin{pgfonlayer}{fg}
        \addplot[draw]
        table[x=c_line_x , y=c_line_y, col sep=semicolon]{figures/ErrorContours/script/cdata/contour_0.001.csv};
        \addplot[draw]
        table[x=c_line_x , y=c_line_y, col sep=semicolon]{figures/ErrorContours/script/cdata/contour_0.0001.csv};
        \addplot[draw]
        table[x=c_line_x , y=c_line_y, col sep=semicolon]{figures/ErrorContours/script/cdata/contour_1e-05.csv};
        \addplot[draw]
        table[x=c_line_x , y=c_line_y, col sep=semicolon]{figures/ErrorContours/script/cdata/contour_1e-06.csv};
        \addplot[draw]
        table[x=c_line_x , y=c_line_y, col sep=semicolon]{figures/ErrorContours/script/cdata/contour_1e-07.csv};
    \end{pgfonlayer}

    \begin{pgfonlayer}{bg}
        \begin{scope}[opacity=1,transparency group]
            \clip (axis cs:1e-5,1e-3) rectangle (1e-2,3e-2);
            \draw[fill=gray!25, draw=none](axis cs:2e-3,3e-2) -- (axis cs:0.6e-6,0.6e-3) -- (axis cs:0.4e-5,0.6e-3) -- (axis cs:1e-2,3e-2) node(foropt){};
        \end{scope}
        \draw[<-](axis cs:2e-4,4.5e-2) -- node[above]{more expensive, same error}  ++ (3,0);
        \draw[->](axis cs:5e-3,1.5e-2) -- ++ (0,-2);
    \end{pgfonlayer}
    \begin{pgfonlayer}{fg}
    \node [above right, xshift=-2.3cm] at (axis cs:  6.068944139550695e-05, 0.027359409978889637) {$\max(e_A, e_B) = 10^{-3}$};
    \node [above right,xshift=0.2cm] at (axis cs:  3.434755181220991e-05, 0.01613664376285744) {$10^{-4}$};
    \node [above right,xshift=0.15cm] at (axis cs:  1.9894938316598997e-05, 0.009153784888423784) {$10^{-5}$};
    \node [above right,xshift=0.15cm] at (axis cs:  1.2889114879713652e-05, 0.005140781873633584) {$10^{-6}$};
    \node [above right,xshift=-0.15cm] at (axis cs:  1.288911487971365e-05, 0.002836777910752072) {$10^{-7}$};
    \end{pgfonlayer}

    \begin{pgfonlayer}{fg}
        \addplot[mark=square*, fill=gray, draw=black, only marks]table[x=time step size Mass-Right, y=time step size Mass-Left, col sep=comma]{figures/ErrorContours/script/sweet.csv};
    \end{pgfonlayer}

\end{loglogaxis}
\end{tikzpicture}
  \caption{Contour plot for varying both time step sizes $\delta t_A$ and $\delta t_B$ of the partitioned oscillator. Contour lines help identify regions of identical error. If one of the ODE integrators dominates the error, the theoretical convergence orders--two for GA and four for RK--can be inferred from the spacing between contour lines along the respective axis. The optimal ratio of time step sizes, where reducing either time step size significantly reduces the error, is highlighted with a gray corridor and gray squares.}
  \label{fig:multirate_oscillator_problem_varyBothDt}
\end{figure}
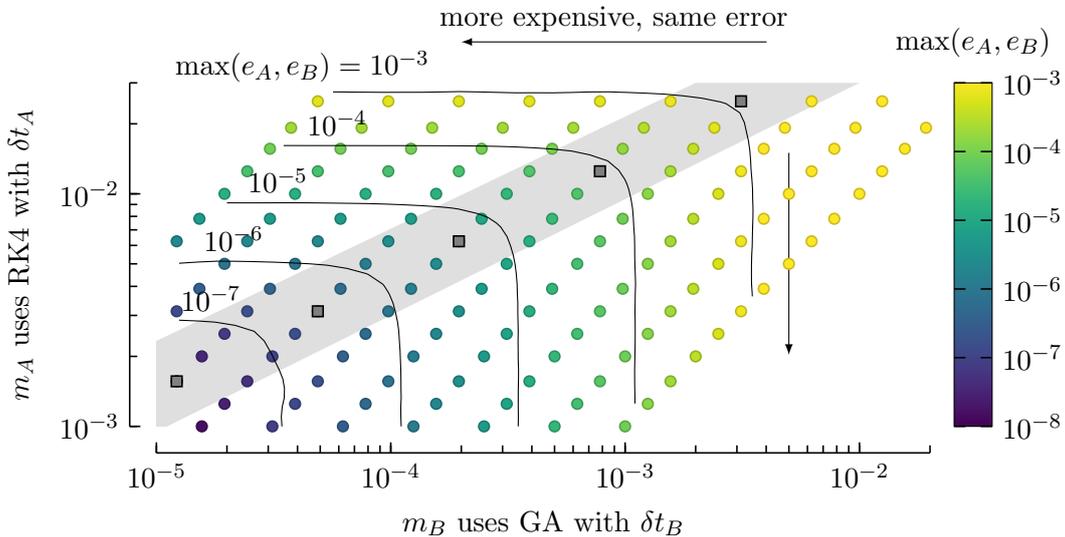

This first experiment already highlights that waveform iteration is a powerful framework for coupling different time integration methods. However, it also shows that balancing the associated errors becomes a non-trivial task. In a second experiment, we use waveform iteration with third-degree B-spline interpolation over a wide range of constant time step and time window sizes to explore the optimal ratio of time step sizes. Specifically, we vary $\Delta t$ from $1/5$ to $1/800$, keeping $A$'s time step fixed as $\delta t_A = \Delta t / 4$, while $B$'s time step $\delta t_B$ is varied from $\delta t_A$ to $\delta t_A / 512$, in order to account for $B$'s lower order (GA) compared to $A$ (RK4). \Cref{fig:multirate_oscillator_problem_varyBothDt} shows a contour plot of the convergence behavior and empirically identifies a corridor of optimal time step ratios (gray squares). As expected, halving $\delta t_A$ requires dividing $\delta t_B$ by four to maintain optimal accuracy.

In a final experiment on the partitioned oscillator, we fix the time step sizes close to the optimal ratio at $\delta_A = 0.005$ and $\delta_B = 0.0002$ (approximately corresponding to the middle gray square in \Cref{fig:multirate_oscillator_problem_varyBothDt}), and increase the time window size from $\Delta t = 0.005$ to $\Delta t = 0.2$, similarly to the thought experiment at the end of \Cref{sec:implementation}. \Cref{tab:oscillator} presents the resulting errors and the average number of coupling iterations per time window. While the error remains stable, the number of iterations increases gradually. For instance, reducing the synchronization frequency by a factor of ten (from $\Delta t = 0.01$ to $\Delta t = 0.1$) results in a 69\% increase in the number of iterations. Repeating the same experiment with quasi-Newton acceleration leads to slightly lower iteration counts, but similar scaling behavior (not shown here, but included in the data publication \cite{Rodenberg2025widata}).

\begin{table}
\centering
\begin{tabular}{cccc}
$\Delta t$ & $e_A$ & $e_B$ & Avg. its. \\
\midrule
0.005 & 7.80E-04 & 7.46E-04 & 3.06 \\
0.01 & 5.78E-06 & 4.93E-06 & 3.08 \\
0.02 & 3.94E-06 & 3.55E-06 & 4.00 \\
0.05 & 4.27E-06 & 3.80E-06 & 4.80 \\
0.1 & 4.39E-06 & 3.89E-06 & 5.20 \\
0.2 & 4.45E-06 & 3.93E-06 & 7.00
\end{tabular}
\caption{Error and average number of iterations for fixed time step sizes $\delta_A=0.005$, $\delta_S=0.0002$ and increasing time window sizes. $\Delta t=0.005$ marks an outlier as $A$ does not have enough samples per time window to build a third-degree B-spline.}
\label{tab:oscillator}
\end{table}

\subsection{Partitioned heat equation}

Next, we evaluate the waveform iteration capabilities of preCICE on a first PDE example, the heat equation,
\begin{equation*}
  \frac{\partial u}{\partial t} = \Delta u + f \;,
\end{equation*}
on the rectangular domain $\Omega=[0,2]\times [0,1]$ and over the time intervall $[0,1]$. We use the manufactured solution \cite[Section 3.1]{Langtangen2017}
\begin{equation*}
    u_\text{exact} (x,y,t) = 1 + (\sin(t)+\cos(t)) x^2 + 3 y^2 + 1.2 t \;,
\end{equation*}
to define initial conditions, Dirichlet boundary conditions on $\partial \Omega$, and the right-hand side $f$.
For the spatial discretization, we use second-order finite elements, which allow exact representation of the polynomial terms $x^2$ and $y^2$ in $u_\text{exact}$, including their fluxes. This lets the spatial error vanish and allows us to focus on the time discretization error. We investigate several Runge-Kutta time-stepping schemes following the approach used in the Firedrake module Irksome \cite{Farrell2021}.

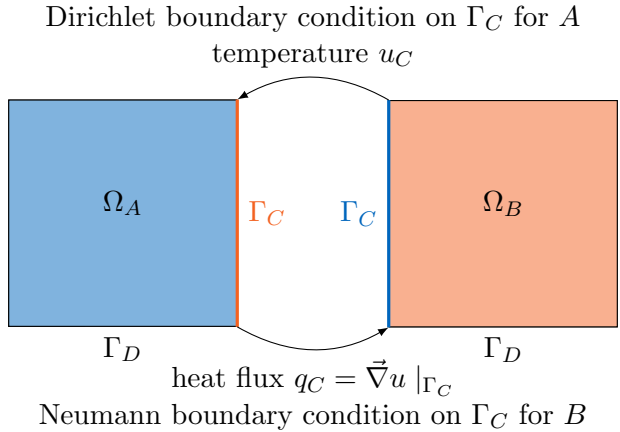
\begin{figure}[t]
  \center
  \begin{tikzpicture}
      \draw[fill=pblue!50](0,0) rectangle ++(1,1);
      \draw
          (0,0) coordinate (a)
          rectangle
          ++(3,3) coordinate (b);
      \node[fill=pblue!50, draw=black, fit=(a) (b),inner sep=0pt] (participantD) {$\Omega_A$};

      \draw
          ($(participantD.south east)+(2,0)$) coordinate (a)
          rectangle
          ++(3,3) coordinate (b);
      \node[fill=porange!50, draw=black, fit=(a) (b),inner sep=0pt] (participantN) {$\Omega_B$};

      \draw[draw=none] (participantD.south east) -- node[below]{$\Gamma_D$} (participantD.south west);
      \draw[draw=none] (participantN.south east) -- node[below]{$\Gamma_D$} (participantN.south west);

      \draw[very thick, porange](participantD.south east) -- node[right]{$\Gamma_C$} (participantD.north east);
      \draw[very thick, pblue](participantN.south west) -- node[left]{$\Gamma_C$} (participantN.north west);

      \draw[->](participantN.north west) to[out=150,in=30] node[above,align=center]{Dirichlet boundary condition on $\Gamma_C$ for $A$\\temperature $u_C$} (participantD.north east);
      \draw[->](participantD.south east) to[out=-30,in=-150] node[below,align=center]{heat flux $q_C = \vec{\nabla} u \mid_{\Gamma_C}$\\Neumann boundary condition on $\Gamma_C$ for $B$} (participantN.south west);
  \end{tikzpicture}
  \caption{Setup of the partitioned heat equation problem. $A$ acts as Dirichlet participant and $B$ as Neumann participant in a non-overlapping Dirichlet-Neumann coupling}
  \label{fig:partitioned-heat-conduction}
\end{figure}

To define a coupled problem, we artificially split the domain into two non-overlapping subdomains, as illustrated in \Cref{fig:partitioned-heat-conduction}. We apply a Dirichlet-Neumann coupling: Participant $A$ uses the interface temperature $u_C$ as a Dirichlet boundary condition and computes the heat flux $q$ on the interface, which is then used by participant $B$ as a Neumann boundary condition. $B$, in turn, returns the interface temperature. The coupling is performed using serial-implicit coupling with reduced quasi-Newton acceleration \cite{Kotarsky2025, Rueth2021}, iterating until the relative residuals of both temperature and flux at the interface fall below $10^{-12}$. Since the meshes match at the coupling interface, no mapping error occurs.
To assess the accuracy of different time-stepping methods, we compute the relative $L^2$-error over the subdomain of $A$ and report the maximum value across all time steps, denoted by $e_A$.
Unlike the oscillator example, this experiment uses a more realistic software setup: We solve both subdomains using FEniCS \cite{Alnaes2015} and use the FEniCS-preCICE adapter \cite{Rodenberg2021} to call preCICE on both sides. The implementation of the time-stepping has been carried out in Vinnitchenko \cite{Vinnitchenko2024}.\footnote{Internally, this approach does not use the interface temperature $u_C$ but its time derivative $\dot{u}_C$ as a Dirichlet boundary condition for the stages of the time stepping schemes \cite{Farrell2021}. Since preCICE does not yet offer an API to obtain the time derivatives of waveforms (see \url{https://github.com/precice/precice/issues/1908}), the Dirichlet participant has to perform an additional reconstruction step to compute the time derivatives.}

\begin{figure}
  \center
  \begin{tikzpicture}[
   mark options={solid}
]
\begin{loglogaxis}[
 tuftelike,
 height=0.5\textwidth,
 width=0.65\textwidth,
 legend cell align=left,
 xlabel={$\delta t_A = \delta t_B = \Delta t / 5$},
 x label style={at={(axis description cs:0.5,-0.1)},anchor=north},
 ylabel={$e_A$},
 y label style={at={(axis description cs:-0.025,0.5)},anchor=south},
 align=center,
 legend style={at={(1.05,0.5)},anchor=west, draw=none, align=left},
 xmin=0.00625,
 ymax=0.1,
 ymin=1e-13,
 xtick = {0.2, 0.1, 0.05, 0.025, 0.0125, 0.00625},
 xticklabels = {$\frac{1}{5}$, $\frac{1}{10}$, $\frac{1}{20}$, $\frac{1}{40}$, $\frac{1}{80}$, $\frac{1}{160}$},
]
    \addplot[dashed, color=red] table[x=time step size Monolithic, y=error Monolithic, col sep=comma]{figures/HigherOrderPartitionedHeat/results/IE_mono.csv};
    \addlegendentry{{IE}, mono}

    \addplot[mark=pentagon, color=red] table[x=time step size Dirichlet, y=error Dirichlet, col sep=comma]{figures/HigherOrderPartitionedHeat/results/IE_3.csv};
    \addlegendentry{{IE}, $p=3$}

    \addplot[dashed, color=pblue] table[x=time step size Monolithic, y=error Monolithic, col sep=comma]{figures/HigherOrderPartitionedHeat/results/GL2_mono.csv};
    \addlegendentry{{GL(2)}, mono}

    \addplot[mark=o, dotted, color=pblue] table[x=time step size Dirichlet, y=error Dirichlet, col sep=comma]{figures/HigherOrderPartitionedHeat/results/GL2_2.csv};
    \addlegendentry{{GL(2)}, $p=2$}

    \addplot[mark=o, color=pblue] table[x=time step size Dirichlet, y=error Dirichlet, col sep=comma]{figures/HigherOrderPartitionedHeat/results/GL2_3.csv};
    \addlegendentry{{GL(2)}, $p=3$}

    \addplot[mark=*,color=pblue] table[x=time step size Dirichlet , y=error Dirichlet, col sep=comma]{figures/HigherOrderPartitionedHeat/results/GL2_5.csv};
    \addlegendentry{{GL(2)}, $p=5$}

   \addplot[dashed, color=gray] table[x=time step size Monolithic, y=error Monolithic, col sep=comma]{figures/HigherOrderPartitionedHeat/results/GL3_mono.csv};
   \addlegendentry{{GL(3)}, mono}

    \addplot[mark=triangle, color=gray] table[x=time step size Dirichlet, y=error Dirichlet, col sep=comma]{figures/HigherOrderPartitionedHeat/results/GL3_3.csv};
    \addlegendentry{{GL(3)}, $p=3$}

    \addplot[mark=triangle*, color=gray] table[x=time step size Dirichlet , y=error Dirichlet, col sep=comma]{figures/HigherOrderPartitionedHeat/results/GL3_5.csv};
    \addlegendentry{{GL(3)}, $p=5$}

    \draw[] ([yshift=+0.2em]axis cs:0.05,0.012) -- node[above]{$\mathcal{O}(\delta t)$} ([yshift=+0.2em]axis cs:0.00625,0.0015);
    \draw[] ([yshift=+0.4em]axis cs:0.05,1.2224e-06) -- node[above]{$\mathcal{O}(\delta t^2)$} ([yshift=+0.4em]axis cs:0.00625,1.91e-8);
    \draw[] ([yshift=+0.3em]axis cs:0.1,6.461244243297452e-07) -- node[above,pos=0.8,xshift=-.1cm]{$\mathcal{O}(\delta t^3)$} ([yshift=+0.3em]axis cs:0.00625,1.577452207836292e-10);
    \draw[] ([yshift=-0.5em]axis cs:0.1,2.3046058578483483e-07) -- node[below,xshift=+.25cm]{$\mathcal{O}(\delta t^4)$} ([yshift=-0.5em]axis cs:0.00625,3.5165494657109807e-12);
    \draw[] ([yshift=-0.4em]axis cs:0.2,1.0241990258728702e-08) -- node[below,xshift=+.35cm]{$\mathcal{O}(\delta t^5)$} ([yshift=-0.4em]axis cs:0.025,3.1256073787624213e-13);

    \end{loglogaxis}
    \end{tikzpicture}
  \caption{Time stepping error for the partitioned heat conduction problem using different time stepping methods. Waveform iteration with B-spline interpolation of degree $p=2,3$, and $5$ is used. The subcycling ratio $\Delta t / \delta t = 5$ is fixed. Monolithic experiments are shown as reference. Similar results can be obtained for the LobattoIIIC scheme with three stages (not shown here, included in the data publication \cite{Rodenberg2025widata}).
  }
  \label{fig:experiment_partitioned_heat_conduction_higher_order}
\end{figure}
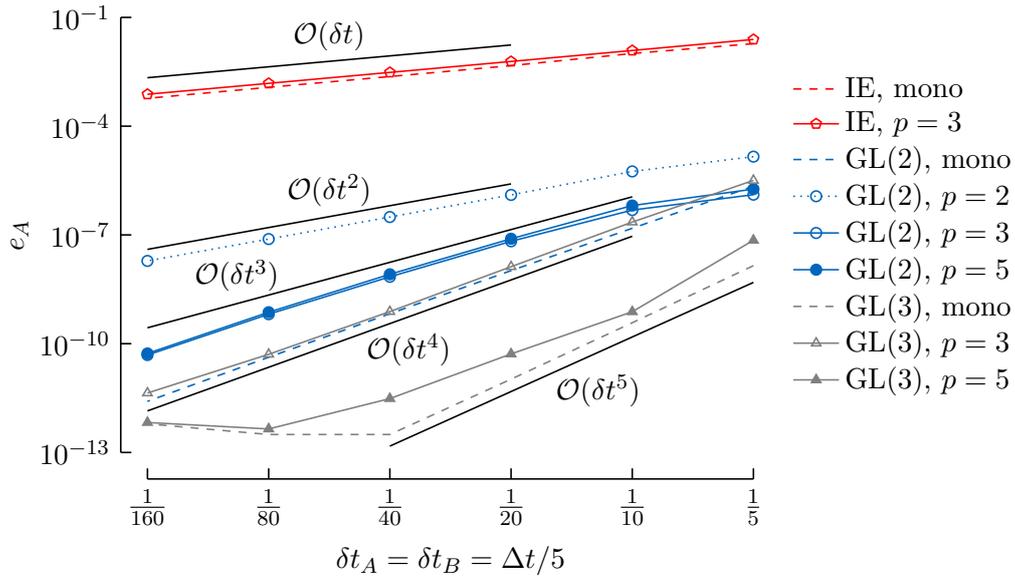

\Cref{fig:experiment_partitioned_heat_conduction_higher_order} presents a convergence study in which we reduce both the time window and time step sizes while keeping the ratio $\Delta t / \delta t=5$ constant, and observe the resulting error $e_A$. All configurations are symmetric: Both participants use the same time stepping methods and identical step sizes. Waveform iteration is performed with B-spline interpolation of degree $p=2,3$, and $5$.

The implicit Euler (IE) method shows first-order convergence, with nearly identical error levels for all $p$. The two-stage Gauss-Legendre method (GL(2)) achieves second-order convergence for $p=2$. For $p=3,5$, the results are almost identical and we observe a convergence order between third order and the expected fourth-order of the monolithic experiment. The three-stage Gauss-Legendre method (GL(3)) achieves only fourth-order convergence for $p=3$. We observe a clear improvement when increasing the degree of the interpolation to $p=5$ and almost reach the expected fifth-order convergence.

Finally, we verify the correct implementation of the quasi-Newton acceleration in the waveform iteration framework by comparing iteration counts to those of the original prototype implementation \cite{Rueth2021}. The results are consistent and successfully reproduced (not shown here; see Rodenberg \cite{Rodenberg2025} for details).

\subsection{Perpendicular flap}

\begin{figure}
  \centering
      \includegraphics[width=.6\textwidth]{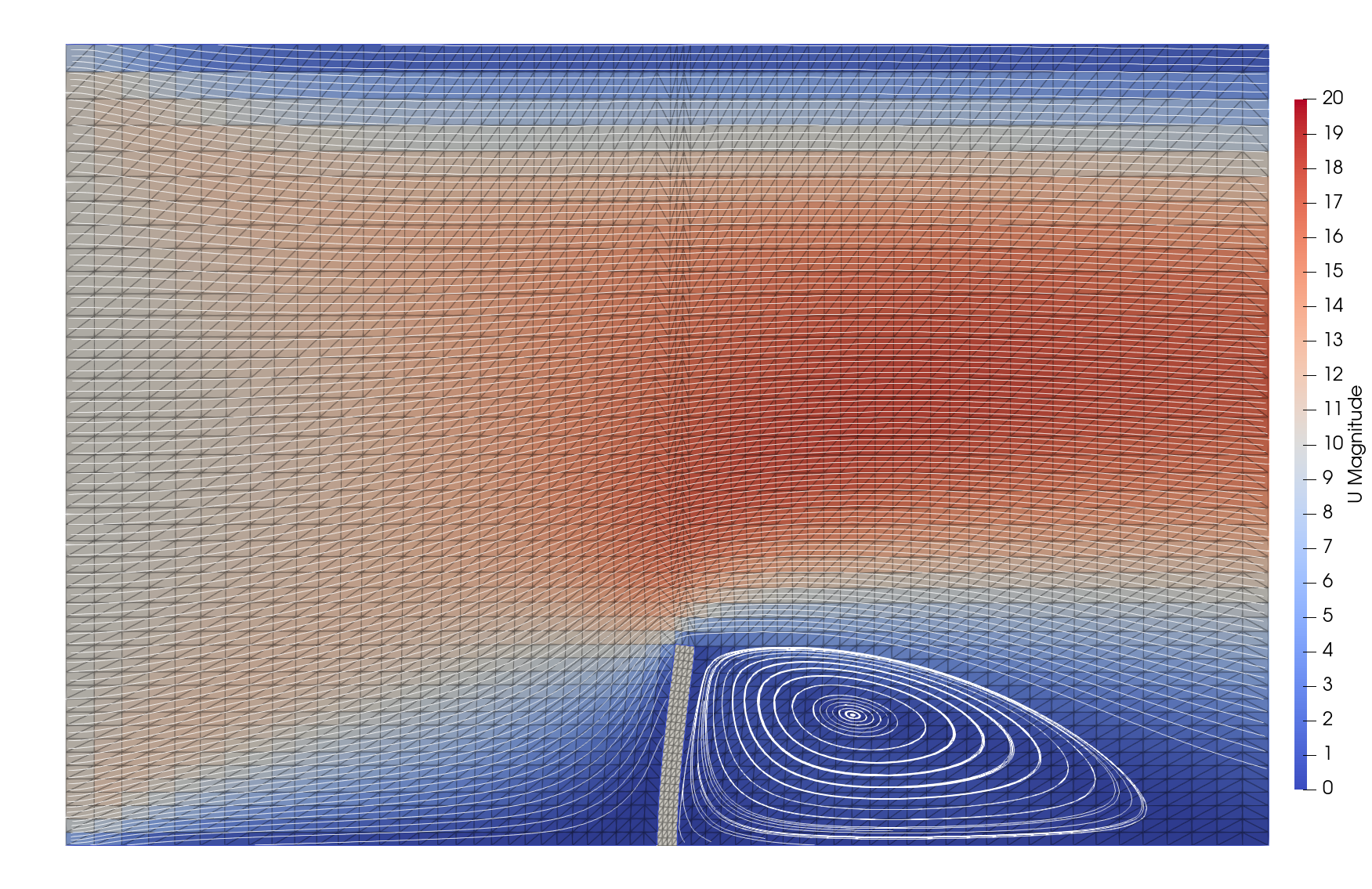}
      \caption{Perpendicular flap with streamlines and discretization mesh of the fluid field and the elastic beam.}
      \label{fig:scenario-perpendicular-flap}
\end{figure}

As a final test case, we consider a classical fluid-structure interaction (FSI) scenario: an elastic flap mounted perpendicularly in a cross-flow, see \Cref{fig:scenario-perpendicular-flap}. The fluid domain is governed by the incompressible Navier-Stokes equations, while the solid domain follows a linear elasticity model. The geometry, boundary conditions, and material parameters follow the setup from the preCICE tutorials\footnote{\url{https://precice.org/tutorials-perpendicular-flap.html}}.

The solid participant uses finite element discretization via FEniCS \cite{Alnaes2015} and is coupled through the FEniCS-preCICE adapter \cite{Rodenberg2021}.
The fluid participant uses finite volume discretization through OpenFOAM \cite{Weller1998_OpenFOAM}, coupled using the OpenFOAM-preCICE adapter \cite{Chourdakis2023}.
We again use a classical Dirichlet-Neumann coupling, where the fluid participants uses a Dirichlet boundary condition at the coupling interface and the solid participant a Neumann boundary condition. We also use again a serial-implicit coupling with reduced quasi-Newton acceleration \cite{Kotarsky2025, Rueth2021}, iterating until the relative residuals of both displacements and forces at the interface fall below $5.0 \cdot 10^{-3}$.
We use non-matching meshes and a partition-of-unity radial-basis-function interpolation \cite{Schneider2025} for data mapping.

For time integration, FEniCS employs the generalized-alpha method.
OpenFOAM also applies a second-order time integration scheme, which, however, effectively degrades to first-order accuracy in the presence of FSI coupling~\cite{Amoros2024}. Achieving true second-order accuracy in a CFD solver under mesh motion and coupling constraints is a non-trivial task and cannot be expected from standard off-the-shelf solvers~\cite{Fehn2021, Rueth2021}.
The coupled system exhibits instability for fluid time step sizes $\delta t_F > 0.01$. This relatively strict time-step limitation on the CFD side, combined with the second-order time integration available on the solid side, makes this setup an interesting candidate for evaluating non-matching time step sizes.

\begin{figure}[h!]
  \begin{tikzpicture}
    \begin{axis}[
        tuftelike,
        xlabel={$t$},
        ylabel={$d_x$},
        align=center,
        width=.45\textwidth,
        height=.35\textwidth,
        xmin=0,
        xmax=5,
        ymin=-0.2,
        ymax=0.4,
        xtick={0,5},
        ytick={-0.2,0,0.2,0.4},
        yticklabel style={
            /pgf/number format/fixed,
            /pgf/number format/precision=2,
        },
        clip=false,
        legend style={at={(1.1,0.5)},anchor=west, draw=none},
        ]
        \addplot[color=black, dotted] table[x=Time, y=Displacement0, col sep=space]{figures/perpflapSubcycling/results/watchpoint_REF.csv};
        \label{plot:REF}
        \addplot[color=red, mark=-, only marks] table[x=Time, y=Displacement0, col sep=space]{figures/perpflapSubcycling/results/watchpoint_MC_S10_rQNWI.csv};
        \label{plot:MC_S10_rQNWI}
        \addplot[color=blue, mark=|, only marks] table[x=Time, y=Displacement0, col sep=space]{figures/perpflapSubcycling/results/watchpoint_SC_S10.csv};
        \label{plot:SC_S10}
        \addplot[color=red, mark=otimes, only marks] table[x=Time, y=Displacement0, col sep=space]{figures/perpflapSubcycling/results/watchpoint_MC_S100_rQNWI.csv};
        \label{plot:MC_S100_rQNWI}
        \addplot[color=blue, mark=oplus, only marks] table[x=Time, y=Displacement0, col sep=space]{figures/perpflapSubcycling/results/watchpoint_SC_S100.csv};
        \label{plot:SC_S100}
        \addplot[color=red, mark=square, only marks, mark size=3] table[x=Time, y=Displacement0, col sep=space]{figures/perpflapSubcycling/results/watchpoint_MC_S500_rQNWI.csv};
        \label{plot:MC_S500_rQNWI}
        \addplot[color=blue, mark=o, only marks, mark size=3] table[x=Time, y=Displacement0, col sep=space]{figures/perpflapSubcycling/results/watchpoint_SC_S500.csv};
        \label{plot:SC_S500}

        \node[anchor=west] at (axis cs:5.5,0.1){
            \small
            \begin{tabular}{r|cc|cc}
                  $\Delta t$ & \multicolumn{2}{c|}{linear (Fig.~\ref{fig:approaches_b})} & \multicolumn{2}{c}{WI (Fig.~\ref{fig:approaches_d})} \\
                  \midrule
                    $0.01$  &\ref{plot:REF}    & 2.01   &\ref{plot:REF}          & 2.01 \\
                    $0.03$  &                  & 2.04   &                        & 2.40 \\
                    $0.05$  &                  & 2.07   &                        & 2.54 \\
                    $0.1$   &\ref{plot:SC_S10} & 2.62   &\ref{plot:MC_S10_rQNWI} & 2.78 \\
                    $0.25$  &                  & 3.75   &                        & 3.90 \\
                    $0.5$   &                  & 4.80   &                        & 5.20 \\
                    $1.0$   &\ref{plot:SC_S100}& 9.60   &\ref{plot:MC_S100_rQNWI}& 6.80 \\
                    $2.5$   &                  & 24.00  &                        & 12.50 \\
                    $5.0$   &\ref{plot:SC_S500}& 8.00   &\ref{plot:MC_S500_rQNWI}& 19.00 \\
            \end{tabular}
        };
    \end{axis}
\end{tikzpicture}
  \caption{Tip displacement (left) and average coupling iterations (right) for the perpendicular flap problem for identical time step sizes of $\delta t = 0.01$ and increasing window sizes. Single-value linear interpolation (linear) is compared to waveform iteration with third-degree B-splines (WI).}
  \label{fig:perpflap_subcycling}
\end{figure}
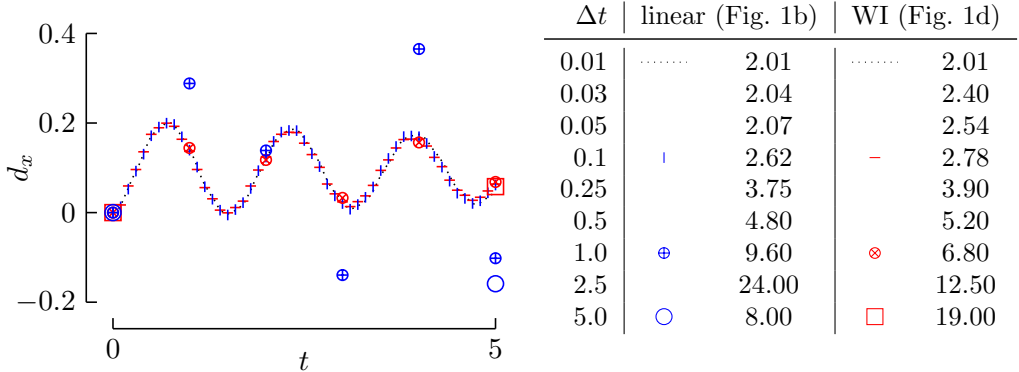

Still, in a first experiment, identical time step sizes are used for both participants, with increasing time window sizes.
We compare waveform iteration with third-degree B-splines (cf.~\Cref{fig:approaches_d}) against the much cheaper single-value linear interpolation (\Cref{fig:approaches_b}), which is often used for FSI problems (e.g., \cite{DeMoerloose2019}).
\Cref{fig:perpflap_subcycling} shows the oscillating tip displacement of the flap over the time interval $[0,5]$ and the average number of coupling iterations for various configurations.

Waveform iteration with third-degree B-splines resolve the tip displacement very well even if we only use a single time window for the complete time interval. Single-value linear interpolation shows good results up to moderate subcycling ($\Delta t = 10 \, \delta t$), but not beyond.
The average number of iterations grows slowly, but not drastically with an increasing window size. For instance, synchronizing only every 10 time steps costs an increase of 38\% in terms of iterations showing the potential. For larger time domains, we even observed a slower increase (not shown).

\begin{figure}[h!]
  \begin{tikzpicture}
    \begin{axis}[
        tuftelike,
        xlabel={$t$},
        ylabel={$d_x$},
        align=center,
        width=.45\textwidth,
        height=.25\textwidth,
        xmin=0,
        xmax=1,
        ymin=0,
        ymax=0.2,
        yticklabel style={
            /pgf/number format/fixed,
            /pgf/number format/precision=2,
        },
        clip=false,
        ytick={0,0.2},
        xtick={0,1},
        legend columns=2,
        legend cell align={left},
        legend style={
            at={(1,0.2)},
            anchor=east,
            draw=none,
            /tikz/column 2/.style={
                column sep=5pt,
            },
        },
        ]
        \addplot[color=black, dotted] table[x=Time, y=Displacement0, col sep=space]{figures/perpflapMultirate/results/watchpoint_REF.csv}; \label{plot:REF}
        \addplot[color=black, mark=o, only marks] table[x=Time, y=Displacement0, col sep=space]{figures/perpflapMultirate/results/watchpoint_MC_20_40.csv}; \label{plot:MR_20_40}
        \addplot[color=brown, mark=-, only marks, mark size=3] table[x=Time, y=Displacement0, col sep=space]{figures/perpflapMultirate/results/watchpoint_MC_20_20.csv}; \label{plot:MR_20_20}
        \addplot[color=blue, mark=|, only marks, mark size=3] table[x=Time, y=Displacement0, col sep=space]{figures/perpflapMultirate/results/watchpoint_MC_20_10.csv}; \label{plot:MR_20_10}
        \addplot[color=red, mark=triangle, only marks, mark size=3] table[x=Time, y=Displacement0, col sep=space]{figures/perpflapMultirate/results/watchpoint_MC_20_5.csv}; \label{plot:MR_20_5}
        \addplot[color=black, mark=star, only marks] table[x=Time, y=Displacement0, col sep=space]{figures/perpflapMultirate/results/watchpoint_MC_20_2.csv}; \label{plot:MR_20_2}

        \node[anchor=west] at (axis cs:1.05,0.1){
            \small
            \begin{tabular}{ccccc}
                                 & $\Delta t$ & $\delta t_F$ & $\delta t_S$ & Avg.~its.
                \\
                \midrule
                \ref{plot:REF}       & $0.02$     & $0.001$ & $0.001$ & 5.34\\
                \ref{plot:MR_20_40}  & $0.2$      & $0.01$  & $0.005$ & 5.00 \\
                \ref{plot:MR_20_20}  & $0.2$      & $0.01$  & $0.01$  & 5.20 \\
                \ref{plot:MR_20_10}  & $0.2$      & $0.01$  & $0.02$  & 5.60 \\
                \ref{plot:MR_20_5}   & $0.2$      & $0.01$  & $0.04$  & 5.80 \\
                \ref{plot:MR_20_2}   & $0.2$      & $0.01$  & $0.1$   & 5.60 \\
            \end{tabular}
        };
    \end{axis}
\end{tikzpicture}
  \caption{Tip displacement (left) and average coupling iterations (right) for the perpendicular flap problem for fixed fluid time step size $\delta t_F = 0.01$, fixed window size $\Delta t = 0.2$, and varying solid time step size $\delta t_S$.}
  \label{fig:perpflap_multirate}
\end{figure}
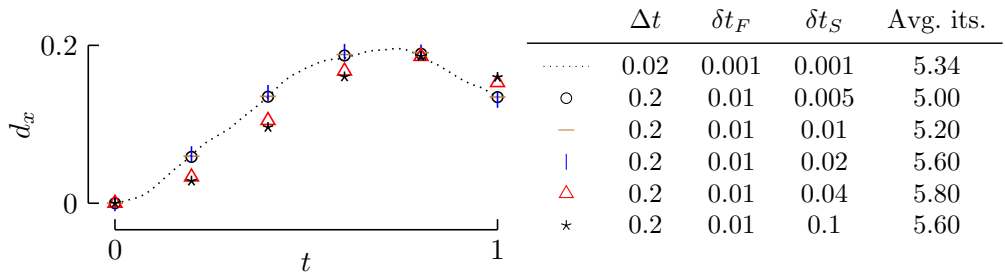

In a second experiment, we fix the fluid time step size at the critical value $\delta t_F = 0.01$ and vary the solid time step size $\delta t_S$. The time window size is kept constant at $\Delta t = 0.2$. As a reference solution, we use a significantly finer simulation with $\delta t_F = \delta t_S = 0.001$ and $\Delta t = 0.02$.
\Cref{fig:perpflap_multirate} shows the tip displacement of the flap over a shorter time interval $[0,1]$, together with the average number of coupling iterations.
We observe that the solid time step size can be increased up to $\delta t_S = 0.02$ without a visible loss in accuracy. For larger step sizes, however, the displacement begins to deviate from the reference solution. Notably, the average number of coupling iterations remains nearly constant across all tested configurations.
Overall, this experiment demonstrates the potential of waveform iteration to independently choose and optimize the time step sizes of the coupled participants and optimize their ratio, improving both flexibility and efficiency in multiphysics simulations.

\section{Conclusions}

We have integrated waveform iteration into the coupling library preCICE and analyzed its efficiency. To the best of our knowledge, this is the first implementation of waveform iteration in a generic PDE coupling software. Achieving this required carefully resolving the interplay with key PDE building blocks, including data mapping, parallel communication, and quasi-Newton acceleration. Despite the technical complexity, only minimal changes are required in existing preCICE adapters, while users gain access to a far more powerful numerical framework.

The new implementation enables coupled participants to use different, even adaptive, time step sizes. This flexibility allows users to balance errors between solvers, though determining optimal step size ratios remains non-trivial. With a sufficiently high interpolation degree, waveform iteration can deliver overall higher-order convergence in time. Importantly, benefits arise even when the individual solvers are only first-order accurate, since users retain the freedom to optimize step sizes independently.

Our experiments also show the potential for drastically fewer synchronization points, an advantage of particular relevance in high-performance computing environments. Quasi-Newton acceleration remains effective, with only a slight increase in iteration counts, even at extreme multirate ratios of up to 100:1 or 1000:1.

Future work will focus on further extending these ratios, for example through data compression strategies, as well as broadening the implementation to support extrapolation, explicit coupling, and interpolation across multiple time windows. We will support the preCICE community in validating the approach on real-world applications.

\section*{Acknowledgments}
The authors gratefully acknowledge the support by the Stuttgart Center for Simulation Science (SimTech).
The authors moreover acknowledge the assistance of AI-tools (ChatGPT-4) for editorial suggestions conformal with the editorial policy of SIAM.

This work was funded by the Deutsche Forschungsgemeinschaft (DFG, German Research Foundation) under Germany's Excellence Strategy, EXC 2075---390740016.

\printbibliography

\end{document}